\documentclass[preprintnumbers, floatfix, letterpaper, superscriptaddress,twocolumn,nofootinbib,altaffilletter,eqsecnum]{revtex4}
\pdfoutput=1
\usepackage{graphicx}
\usepackage{amsmath}
\usepackage{subfigure}
\usepackage{amssymb}
\usepackage{amsfonts}
\usepackage{mathtools}
\usepackage{amssymb}
\usepackage{enumerate}
\usepackage{xcolor}
\usepackage{bm}
\usepackage{mathrsfs}
\usepackage{epstopdf}
\usepackage{url}
\usepackage{footnote}
\usepackage{textcomp}
\usepackage{dsfont}
\usepackage{hyperref}
\usepackage{enumerate}   
\usepackage{appendix}
\usepackage{textcomp}
\usepackage{tipa}

\newcommand{\bea}{\begin{eqnarray}}
\newcommand{\eea}{\end{eqnarray}}
\newcommand{\ba}{\begin{align}}
\newcommand{\ea}{\end{align}}

\makeatletter
\newcommand*{\rom}[1]{\expandafter\@slowromancap\romannumeral #1@}
\makeatother

\begin{document}

\preprint{RIKEN-iTHEMS-Report-25, NITEP 263}

\title{Shadow of rotating black holes with consistent thermodynamics}

\author{Che-Yu Chen}
\email{b97202056@gmail.com}
\affiliation{RIKEN iTHEMS, Wako, Saitama 351-0198, Japan}

\author{Chiang-Mei Chen} \email{cmchen@phy.ncu.edu.tw}
\affiliation{Department of Physics, National Central University, Zhongli, Taoyuan 320317, Taiwan}
\affiliation{Center for High Energy and High Field Physics (CHiP), National Central University, Zhongli, Taoyuan 320317, Taiwan}
\affiliation{Asia Pacific Center for Theoretical Physics (APCTP), Pohang 37673, Korea}

\author{Nobuyoshi Ohta} \email{ohtan.gm@gmail.com}
\affiliation{Nambu Yoichiro Institute of Theoretical and Experimental Physics (NITEP), Osaka Metropolitan University, Osaka 558-5585, Japan}
\affiliation{Asia Pacific Center for Theoretical Physics (APCTP), Pohang 37673, Korea}


\begin{abstract}

Quantum effects in general induce scale dependence in the coupling constants. We explore this possibility in gravity, with a scale-dependent Newton coupling. When applied to Kerr black holes with such a running coupling, the consistency of black hole thermodynamics requires that the Newton coupling have a specific dependence on the black hole parameters. In this work, we consider such a class of Newton couplings and look for the possible observational implications on the highly lensed images of the black holes. In addition to placing constraints on the parameter space of the model through the latest Sgr A* images, we find that the variations in the shape of shadows in a large portion of the parameter space can be qualitatively captured by a quantity solely defined by the event horizon. Most importantly, the consistency of thermodynamics suggests a lower bound on the shadow size, beyond which either horizon disappears, or the shadow cannot keep the standard D-shaped structure. The possibility that the black holes in this model could spin faster than the Kerr bound, and the physical implications of the resulting cuspy shadows, are also discussed.  
\end{abstract}

\maketitle


\section{Introduction}

Black hole solutions represent fundamental spacetime configurations in Einstein's theory of General Relativity (GR). However, the presence of singularities signals the limitations of GR. Indeed, there have been a lot of proposals to modify the metric to resolve singularities in the black hole geometry~\cite{1968qtr..conf...87B, Dymnikova:1992ux, Hayward:2005gi, Bambi:2013ufa}. For reviews and more references, see Refs.~\cite{Torres:2022twv,Lan:2023cvz}.
The resolution of the problem comes most probably from the quantum effects, which might produce such modifications of the metric.
In the study of black holes, the associated thermodynamic properties may offer a remarkable window into quantum effects manifesting at the semiclassical level.

Among others, one of the promising approaches to quantum gravity within the framework of quantum field theory is to use the functional renormalization group (FRG). In this approach, called asymptotic safety~\cite{1979grec.conf..790W}, we search for nontrivial ultraviolet (UV) fixed points of the couplings in the theory. If a quantum field theory can be consistently restricted to a finite set of operators with well-defined UV fixed points under the renormalization group flow, the theory becomes scale invariant at high energies, and the UV divergences may be under control. A significant amount of evidence has been obtained in favor of this scenario~\cite{Reuter:1996cp, Souma:1999at}. For recent reviews on this subject, see Refs.~\cite{Percacci:2017fkn, Eichhorn:2018yfc,2019qgfr.book.....R}.

One of the important consequences in the asymptotic safety is that the couplings of the theory become functions of an energy scale. 
In the search for physical implications of this energy scale for the black hole solutions, its identification with some length scale has been considered in the geometry of black holes~\cite{Bonanno:2000ep,Bonanno:2006eu,Reuter:2004nv,Falls:2010he,Reuter:2010xb,Harst:2011zx,Falls:2012nd,Koch:2013owa,Litim:2013gga,Koch:2014cqa,Bonanno:2017zen,Pawlowski:2018swz,Platania:2020lqb,Ishibashi:2021kmf,Ruiz:2021qfp,Lu:2024ppa}. The identification generally depends on the 
spacetime coordinates and associated physical quantities. However, any fundamental principle that uniquely determines this identification had not been found, other than the dimensional reason.
On the other hand, when the rotating Kerr black hole is considered, it is found that the first law of black hole
thermodynamics is not satisfied straightforwardly~\cite{Reuter:2010xb}. Since the first law of thermodynamics is the fundamental law of
energy conservation, it must be valid not only semiclassically but also quantum mechanically. It has been shown that the consistency condition of the first law of the black hole thermodynamics imposes a very general, albeit not so strong, nontrivial constraint on the dependence of the Newton coupling on the black hole parameters and geometry in the quantum theory~\cite{Chen:2022xjk, Chen:2023wdg}.

We emphasize that this constraint is valid beyond asymptotic safety. As mentioned above, in the attempt to resolve singularity in the black holes, modifications of the geometry have been considered, some of which may be interpreted as scale dependence of the Newton coupling. However, it is quite often not very clear what principle governs the modification apart from the singularity resolution. Including such cases, if the Newton coupling gets any dependence on the geometry in any attempt to modify the geometry of the rotating black holes, the above constraint must be obeyed in order to have consistent thermodynamics.
It is then important to study what are the observational implications of the dependence on the black hole geometry of the Newton coupling, and ask if we can test them and/or impose any constraint on the dependence on the black hole parameters through black hole observations.
This subject gets its urgency if we recall that almost all the black holes in our Universe would be rotating ones, which have to satisfy the above consistency condition.

A class of black hole observations that is suitable for our purpose is the shadow images cast by black holes \cite{Falcke:1999pj,EventHorizonTelescope:2019dse,EventHorizonTelescope:2022wkp,Chen:2022scf}. Because of the strong gravitational fields near black holes, the light rays that propagate around the black hole can be highly lensed and orbit the black hole multiple times, as long as their trajectories get close enough to the black hole. The black hole images that appear on the observer's image plane consist of the higher-order images contributed by the highly lensed light rays, on top of the direct emission contributed by the light rays that propagate directly to the observer, without orbiting the black hole \cite{Bozza:2010xqn,Gralla:2019xty,Johnson:2019ljv}. The more highly lensed light rays, as well as their higher-order images, are expected to carry a larger fraction of information about the black hole geometry itself, and are thus less sensitive to the emission profiles under consideration \cite{Cunha:2018acu}. In particular, as one increases the order of lensed images, the higher-order images quickly converge to a closed contour that we term ``critical curve"{\footnote{This terminology is adopted from Ref.~\cite{Gralla:2019drh}.}} on the image plane \cite{Gralla:2019drh}, which corresponds to the impact parameter of infinitely lensed light rays orbiting on a set of unstable spherical orbits around the black hole. This set of spherical orbits and the shadow critical curve on the images are solely determined by the black hole metric through the photon geodesic equations. Therefore, shadow critical curves carry pure information about the black hole geometry and can be very crucial to probe the black hole geometry itself in a theory-agnostic manner. For nonrotating black holes, it is well known that the shadow critical curve appears as a perfect circle, while a rotating black hole, e.g., the Kerr one, would cast a D-shaped shadow due to the frame-dragging effects \cite{Luminet:1979nyg}. By properly identifying the observables that describe the size or the shape of the critical curve, one could use them to constrain the parameter space of the model or extract relevant information about the black hole geometry \cite{Johannsen:2015hib,Younsi:2021dxe}. This line of research has been widely considered in the literature \cite{Wielgus:2021peu,Vagnozzi:2022moj}, for instance, for black holes in modified theories of gravity \cite{Zhu:2019ura,Khodadi:2020gns,Ghosh:2020spb,Ghosh:2022kit,Jusufi:2022loj,EventHorizonTelescope:2021dqv,Afrin:2021wlj,Papnoi:2021rvw,Pantig:2022ely,Kuang:2022ojj,Khodadi:2022pqh,Meng:2022kjs,Khodadi:2024ubi}, black holes with quantum corrections \cite{Brahma:2020eos,Eichhorn:2021iwq,KumarWalia:2022ddq,Afrin:2022ztr,Islam:2022wck,Zhang:2023okw,Jiang:2023img,Cao:2024vtq,Contreras:2019cmf,Koch:2025gaw}, horizonless compact objects \cite{Dey:2020bgo,Bouhmadi-Lopez:2021zwt,Shaikh:2021cvl,Rahaman:2021web,Shaikh:2022ivr,Olmo:2023lil,Nguyen:2023clb}, and several parametrizations of non-Kerr spacetimes \cite{Ghasemi-Nodehi:2015raa,Younsi:2016azx,Wang:2017hjl,Shaikh:2019fpu,EventHorizonTelescope:2020qrl,Chen:2022lct,Konoplya:2021slg}.

As the first step toward testing the consistency of black hole thermodynamics through black hole observations, in this paper, we consider a class of minimal extensions of the Kerr spacetime. The metric of this model, which was proposed in Ref.~\cite{Chen:2023wdg}, has the form of the usual Kerr one, but the Newton coupling takes the form  $G(r,M,a)=G(r^2+a^2,Mr)$, replacing the original Newton coupling constant $G_0$. As we will show in Sec.~\ref{sec:thermo}, this class of minimal extensions ensures the integrability of the first law of black hole thermodynamics and hence the consistency condition. However, depending on the explicit parametrization of the Newton coupling under consideration, the model can still have a huge range of parameter space. In this paper, we take a general parametrization of the Newton coupling $G(r^2+a^2,Mr)$, consider a few representative cases, and perform a thorough analysis of how the shape and size of shadows vary in a wide range of parameter space. Our analysis is twofold: First, we would like to place constraints on the parameter space of the model. This can be achieved by using the available constraint on the observed size of Sgr A* shadow images \cite{EventHorizonTelescope:2022xqj}, incorporated by the accurate measurements on the mass-to-distance ratio of Sgr A* using surrounding stellar dynamics. Second, a crucial but much more difficult task of this work is to identify possible observational features or implications that can be shared within a sufficiently large (if not all) parameter space of this model. This direction is important in order to draw conclusions regarding the observational implications of the consistency of black hole thermodynamics in general. In particular, if such features or implications exist, they can assist in the attempt to test the consistency conditions of black hole thermodynamics through shadow observations.

The paper is organized as follows. In Sec.~\ref{sec:thermo}, we review the black hole thermodynamics and the conditions for the black hole metric in order to satisfy the consistency of thermodynamics, from which we introduce the class of minimal extensions of Kerr black holes we will be considering. Then, we introduce our theoretical setup in Sec.~\ref{sec:th}. In Sec.~\ref{sec:geodesicshadow}, we briefly review the null geodesic equations, the spherical photon orbits, and the derivation of shadow critical curves. Then, in Sec.~\ref{subsec:obs}, we introduce two shadow observables, which quantify the apparent size and the distortion of the critical curves, respectively. In Sec.~\ref{sec:results}, we present our results on the black hole shadows in our model. In particular, we identify possible observational implications on the shadow critical curves that could be useful in terms of testing the consistency of black hole thermodynamics through black hole images. Then, in Sec.~\ref{sec:superspin}, we discuss the possibility of going beyond the Kerr bound on the black hole spin in our model, and reveal an interesting shadow feature in this case. Finally, we conclude in Sec.~\ref{sec:con}.

\section{Consistency of thermodynamics}
\label{sec:thermo}

In this section, we review how the consistency of the first law of the thermodynamics of the rotating Kerr black hole constrains the dependence of the Newton gravitational coupling on the black hole parameters~\cite{Chen:2022xjk, Chen:2023wdg}.

The metric of the rotating Kerr black hole is given by
\begin{align} \label{ksol}
ds^2 = &- \frac{\Delta}{\Sigma} \left( dt - a \sin^2\theta\, d\varphi
\right)^2 + \frac{\Sigma}{\Delta}\, dr^2 \nonumber\\&+ \Sigma\, d\theta^2 +
\frac{\sin^2\theta}{\Sigma} \left[ a \, dt - (r^2 + a^2)\, d\varphi
\right]^2\,,
\end{align}
where
\begin{eqnarray}
\Delta = r^2 + a^2 - 2 G M r, \qquad \Sigma = r^2 + a^2 \cos^2\theta\,.
\label{Kerrbh}
\end{eqnarray}
Here, $M$ is the mass of the black hole, $a$ is the black hole's specific angular momentum parameter
(angular momentum per unit mass), and $G=G_0$ is the Newton coupling constant.

It is well known that for the Kerr spacetime the first law of black hole thermodynamics holds:
\begin{eqnarray}
dE = T \, dS + \Omega \, dJ\,,
\label{first}
\end{eqnarray}
where $S$ and $T$ denote the entropy and temperature of the black
hole, respectively. The thermodynamic quantities are related as
\begin{eqnarray}
E = M\,, \qquad J = M a\,, \qquad \Omega = \frac{a}{r_h^2 + a^2}\,,\label{termoquantity}
\end{eqnarray}
where $r_h$ is the radius of the outer horizon, determined by the
condition
\begin{eqnarray}
\Delta(r_h) = r_h^2 + a^2 - 2 G M r_h = 0\,.
\label{horizon}
\end{eqnarray}
In this thermodynamic framework, $M$ and
$a$ are treated as independent variables. According to the first law, the differential of the entropy can be expressed as
\begin{equation}
dS(M, a) = \frac{1 - a \Omega}{T} dM - \frac{M \Omega}{T} da =
\partial_M S \, dM + \partial_a S \, da\,.
\end{equation}
To ensure internal consistency, the mixed partial derivatives of $S$
must satisfy the integrability condition,
\begin{equation}
\partial_a \partial_M S = \partial_M \partial_a S\,,
\label{consistency}
\end{equation}
for the $\partial_M S$ and $\partial_a S$ derived from the first law.

If the Newton coupling is a constant with the horizon radius $r_h$ determined from \eqref{horizon}, the condition~\eqref{consistency} is obviously satisfied.
However, when the Newton coupling becomes scale dependent due to, for instance, quantum effects, i.e., instead of $G=G_0$ a constant, when $G = G(r, M, a)$, this condition imposes nontrivial constraint on $G$. To see this, let us first note that the temperature is computed via the surface gravity $\kappa$, giving
\begin{equation}
T = \frac{\kappa}{2\pi} = \frac{r_h - M G(r_h, M, a) - M r_h
\partial_{r_h} G(r_h, M, a)}{2 \pi (r_h^2 + a^2)}\,.
\end{equation}
In addition, on the horizon, the condition~\eqref{horizon} implicitly defines{\footnote{For an asymptotically flat black hole with the ADM mass given by $M$ and the Newton coupling going to the finite constant $G_0$ in the asymptotic region, the energy $E$ and the horizon angular velocity $\Omega\equiv |g_{t\varphi}|/g_{\varphi\varphi}$ on the horizon $r_h$ are still given by Eq.~\eqref{termoquantity}.}} $r_h = r_h(M, a)$.
It is then helpful to express all relevant derivatives, $\partial_M, \partial_a, \partial_{r_h}$, and use the chain rule to relate them. In the analysis in Ref.~\cite{Chen:2023wdg}, it is shown that the integrability condition~\eqref{consistency} for the first law leads to
\begin{eqnarray}
M a \, \partial_M G + r_h^2 \, \partial_a G - a r_h \, \partial_{r_h} G = 0\,.
\end{eqnarray}
This constraint must be satisfied in order for the running Newton coupling to fulfill the thermodynamic consistency. A general solution to this condition is
\begin{equation}
G(r_h, M, a) = G(r_h^2 + a^2, M r_h)\,.
\label{gd}
\end{equation}
Note that $r_h^2 + a^2$ is proportional to the horizon area,
specifically $A = 4 \pi (r_h^2 + a^2)$. Thus, the Newton coupling must
depend on the horizon area and the product $M r_h$. The entropy $S$ can be computed from the following integration
\begin{equation}
S=\int\frac{dA}{4\left(G+M\partial_M G\right)}\,,
\end{equation}
where the denominator in the integrand is a function of $A$ because of Eqs.~\eqref{horizon} and \eqref{gd}, ensuring the integrability of $S$.

Strictly speaking, the relation \eqref{gd} holds on the horizon. However, it is
natural from the viewpoint of continuity to assume an extension of this form beyond the horizon, namely $G(r, M, a) = G(r^2 + a^2, M r)$. In
Ref.~\cite{Chen:2023wdg}, this assumption is used to propose a desirable identification for resolving the ring singularity in rotating black holes.

\section{Setup}
\label{sec:th}

Based on the discussion in the previous section, we consider the rotating black hole spacetime described by the line element~\eqref{ksol}, with the original constant Newton coupling $G_0$ replaced by a running Newton coupling $G(r,M,a)$. Depending on the explicit form of $G(r,M,a)$, there may be one or multiple horizons, which can be determined by solving the equation $\Delta(r_h) = 0$. The outer event horizon $r_h$, which is relevant in the context of black hole thermodynamics, corresponds to the largest real solution of the equation $\Delta (r_h)= 0$.

For later convenience, we define the dimensionless parameter
\begin{equation}
\epsilon = 1 - \frac{a^2}{r_h^2 \left[ 1 - 2 M G'(r_h) \right]}\,, 
\label{defepsilon}
\end{equation}
where the prime denotes the derivative with respect to $r$, and here we explicitly show only the $r$-dependence of $G$ for simplicity. The physical interpretation of this parameter can be understood as follows. As we mentioned, the horizon is determined by $\Delta=0$, i.e., Eq.~\eqref{horizon}. If there are multiple horizons, the extremal condition occurs when the horizons are degenerate:
\bea
\Delta'(r_h) =2 r_h -2 G'(r_h)Mr_h-2 G(r_h)M=0\,.
\label{extremal}
\eea 
The extremal configuration exists only if there is a real solution to Eq.~\eqref{extremal} [and Eq.~\eqref{horizon}]. Such a configuration can be obtained by solving Eqs.~\eqref{horizon} and \eqref{extremal}. One gets
\bea
r_h^2 - a^2 - 2 G'(r_h) M r_h^2 = 0\,,
\label{cond_extremal}
\eea
which is the extremal condition for $r_h$. Comparing Eq.~\eqref{cond_extremal} and the definition of $\epsilon$, one can see that $\epsilon$ vanishes at the extremal configuration when $a\ne0$. For instance, when $G=G_0$ is a constant, $\epsilon=0$ when $r_h=|a|$, which is the extremal limit for the Kerr black hole. Therefore, one can interpret $\epsilon$ as the ``deviation parameter from extremality'' (hereafter referred to as extremality deviation) that quantifies how different the metric with a given parameter choice is from the extremal configuration. In addition, when $a=0$, we always have $\epsilon=1$. Later, we will demonstrate that, in the model we are considering, the parameter $\epsilon$ could be qualitatively read off from the black hole shadows, although $\epsilon$ is purely defined by the event horizon $r_h$ while the shadows are observables defined by spherical photon orbits outside the horizon. In addition, although $\epsilon$ is interpreted as the extremality deviation, its operational definition~\eqref{defepsilon} only requires the radius of the outer event horizon $r_h$. This means that even in the scenario where the spacetime has only one event horizon, one can still calculate $\epsilon$. We will come back to this point in Sec.~\ref{sec:results}.

We have seen in Sec.~\ref{sec:thermo} that if we take into account the consistency of thermodynamics when going beyond the Kerr geometry, one can promote the Newton constant $G_0$ to the running Newton coupling $G(r, M, a)$ of the form~\cite{Chen:2022xjk, Chen:2023wdg}
\begin{equation}
G(r,M,a) = G\left( G_0/(r^2 + a^2), M r \right) \,. \label{generalgfunction}
\end{equation}
In the natural unit $c=\hbar=1$ that we use, $[\textrm{Time}] = [\textrm{Length}] = 1/[\textrm{Mass}]$ and both the terms $G_0/(r^2+a^2)$ and $Mr$ are dimensionless. However, this unit obscures where the quantum effects are. It is not difficult to recover the dependence on $\hbar$ and $c$, and we find
\bea
\frac{G_0}{r^2+a^2} \Rightarrow  \frac{\hbar\, G_0}{(r^2+a^2/c^2)c^3}\,;\qquad
\frac{1}{Mr} \Rightarrow\frac{\hbar}{Mrc}\,.\label{fullunit}
\eea
Thus we find that the scale dependence indeed disappears if we take the classical limit $\hbar\to 0$. That being said, let us use the natural unit again for simplicity of presentation.

A convenient way of parametrizing $G(r,M,a)$ is to introduce the variable
\begin{equation}
z \equiv \left( \frac{G_0}{r^2 + a^2} \right)^l \frac{1}{\left( M r \right)^p} \,, \label{zgeneral}
\end{equation}
where $l$ and $p$ are integers. Then, the Newton coupling is assumed to be a function of $z$, i.e., $G(z)$. We also assume that the asymptotic region $r\rightarrow\infty$ corresponds to the limit $z\rightarrow0$. This means $2l+p\ge 1$. In particular, assuming that the Newton coupling $G$ reduces to $G_0$ at $r\rightarrow\infty$, we can consider the function of the form
\bea
G(z) \equiv G_0 f(z) \,, \qquad \textrm{with} \quad f(0) = 1\,.
\eea
The point is that the dependence of $G(z)$ on the mass $M$ and the spin $a$ can only appear in $z=z(M,a)$ through Eq.~\eqref{zgeneral}. Other parameters that may be interpreted as quantum corrections or modified gravity effects can appear in $G(z)$ only in a manner that they do not depend on $(M,a)$ in order to fulfill the consistency condition of black hole thermodynamics.

In this work, we will consider four specific choices of $(l,p)$, i.e., $(l,p)=(1,-1)$, $(1,0)$, $(1,4)$, and $(0,1)$. A few remarks about these choices of $(l,p)$ are enumerated in the following:
\begin{itemize}
\item First, the choice of $(l,p)=(1,-1)$ belongs to a special case from the perspective of dimensional analysis. Define a set of dimensionless quantities{\footnote{The standard Kerr bound is given by $|a_*|\le1$.}}
\begin{equation}
r_*\equiv\frac{r}{G_0M}\,,\quad a_*\equiv\frac{a}{G_0M}\,.
\end{equation}
Then the variable $z$ given in Eq.~\eqref{zgeneral} can be rewritten as
\begin{equation}
z=\frac{1}{\left(G_0M^2\right)^{p+l}\left(r_*^2+a_*^2\right)^l r_*^p}\,.\label{zgeneralrescale}
\end{equation}
One can see that when $p+l>0$ and for a given $a_*$, the contributions from the running of $G$ through the dependence on $z$ will be suppressed by the black hole mass $M$, i.e., $z\propto1/M^{2p+2l}$. After restoring $\hbar$ in the expression, it becomes clear that this corresponds to the suppression of quantum effects by the black hole mass at the horizon scale, i.e., $z\propto \left(M_p/M\right)^{2p+2l}$, with $M_p$ the Planck mass. For supermassive black holes, the contributions from the running of $G$ can then be unmeasurably small~\cite{Glampedakis:2021oie}. However, if $(l,p)=(1,-1)$, or in general $p=-l$, such a mass suppression does not occur.{\footnote{Actually, this is precisely the combination that $\hbar$ drops out. However, as emphasized in the Introduction, we can consider the scale dependence in a broader context with thermodynamic consistency.}} In this case, all the coupling constants that appear in $G(z)$ can be directly constrained through black hole observations, irrespective of the mass of the black holes being observed.

\item If $p=0$, the dependence of $z$ on the mass $M$ in Eq.~\eqref{zgeneral} disappears. Specifically, $(l,p)=(1,0)$ belongs to the mass-independent scale identification in the FRG method, which was discussed in Ref.~\cite{Chen:2022xjk}.

\item In Ref.~\cite{Chen:2023wdg}, the authors focused on $l=1$ and constructed a class of nonsingular black hole spacetimes, which requires $p > 3$, based on the concept of asymptotic safety by taking
\begin{equation}
G(z) = \frac{G_0}{1 + \tilde{\omega} z}\,,
\end{equation}
where $\tilde\omega>0$ is a coupling constant. At spatial infinity $r\rightarrow\infty$, we have $z\rightarrow0$ and $G(z)\rightarrow G_0$. On the other hand, near the origin $r=0$, we have $z\rightarrow\infty$ and $G(z)$ approaches zero, hence the singularity could be resolved. In this work, we will consider $(l,p)=(1,4)$ as an example inspired by this type of nonsingular black hole models.

\item The last case $(l,p)=(0,1)$ represents the spin-independent scale identification. 
\end{itemize}

By introducing a proper parameterization of $G(z)$, we expect that the analysis of the aforementioned choices of $(l,p)$ should be able to cover the qualitative feature of all the possible combinations of $G\left(G_0/(r^2+a^2),Mr\right)$ that satisfy the asymptotic flatness condition.

\section{Null geodesic and shadow observables}\label{sec:geodesicshadow}

The main aim of this paper is to investigate the shadows cast by the black holes that satisfy the thermodynamic consistency. In what follows, we will focus on the black hole metric given by Eq.~\eqref{ksol}, with a running Newton coupling \eqref{generalgfunction}. In particular, we will study the shadow critical curve on the image plane as seen by a distant observer. Essentially, the critical curve is defined by the impact parameters of the unstable spherical photon orbits around the black hole. These orbits are unstable in the sense that if one slightly perturbs the photons moving on them, these photons would either escape away from the black hole or fall into the horizon. In particular, they represent the asymptotic limit of orbits on which light rays are infinitely lensed. Most importantly, they are solely determined by the black hole spacetime itself and do not depend on the emission models or other astrophysical environments around the black hole. In this section, we will first review the null geodesic equations for the spacetime \eqref{ksol} and derive the equations that determine the shadow critical curves (see also Ref.~\cite{Tsukamoto:2017fxq}). Then, we will introduce the observables that quantify the size and shape of the shadow critical curves.     

\subsection{Shadow critical curves}

It turns out that the geodesic equations for the spacetime \eqref{ksol} with a running $G(r,M,a)$ are separable because of the existence of a Carter-like constant. This can be demonstrated using the Hamilton-Jacobi approach. Let us consider the Lagrangian
\begin{equation}
\mathcal{L}=\frac{1}{2}g_{\mu\nu}\dot{x}^\mu\dot{x}^\nu\,,\label{lagrangiangeo}
\end{equation}
where the dot represents the derivative with respect to the affine parameter $\lambda$. Because of the axisymmetry and stationarity of the spacetime \eqref{ksol}, there are two conserved quantities along the geodesics $E = - p_t$ and $L = p_\varphi$, which correspond to the energy and the azimuthal angular momentum on the geodesic. The conjugate momenta $p_\mu\equiv\partial\mathcal{L}/\partial\dot{x}^\mu$ are defined in the standard way. One can then invert $E$ and $L$ to get the expressions for $\dot{t}$ and $\dot{\varphi}$:
\begin{eqnarray}
\Sigma\dot{t} &=& a \left( L - a E \sin^2\theta \right) + \frac{r^2 + a^2}{\Delta} \left[ (r^2 + a^2) E - a L \right]\,, 
\nonumber\\
\Sigma\dot{\varphi} &=& \frac{L}{\sin^2\theta} - a E + \frac{a}{\Delta} \left[ (r^2 + a^2) E -a L \right]\,.\label{tphidot}
\end{eqnarray}

The Hamiltonian $\mathcal{H}$ associated with the Lagrangian \eqref{lagrangiangeo} is given by
\begin{equation}
\mathcal{H}=\frac{1}{2}p_\mu p^\mu\,.
\end{equation}
Therefore, the Hamilton-Jacobi equation $\partial\mathcal{S}/\partial\lambda+\mathcal{H}=0$ can be written as
\begin{equation}
\frac{\partial\mathcal{S}}{\partial\lambda} = -\frac{1}{2} g^{\mu\nu} \frac{\partial\mathcal{S}}{\partial x^\mu} \frac{\partial\mathcal{S}}{\partial x^\nu}\,,\label{HJeq}
\end{equation}
where $\mathcal{S}$ is the Jacobi action. It turns out that the metric \eqref{ksol} allows for a separable $\mathcal{S}$ as a solution to Eq.~\eqref{HJeq}:
\begin{equation}
\mathcal{S}=-Et+L\varphi+\mathcal{S}_r(r)+\mathcal{S}_\theta(\theta)\,.\label{Sant}
\end{equation}
Inserting Eq.~\eqref{Sant} into Eq.~\eqref{HJeq}, we get
\begin{align}
-\frac{\left[\left(r^2+a^2\right)E-aL\right]^2}{\Delta}+\left(L-aE\right)^2+\Delta\left(\frac{d\mathcal{S}_r}{dr}\right)^2\nonumber\\=\cos^2\theta\left(a^2E^2-\frac{L^2}{\sin^2\theta}\right)-\left(\frac{d\mathcal{S}_\theta}{d\theta}\right)^2\,.
\end{align}
The left-hand side of this equation depends only on $r$, and the right-hand side only depends on $\theta$. Therefore, the functions $d\mathcal{S}_r/dr$ and $d\mathcal{S}_\theta/d\theta$, which relate to $\dot{r}$ and $\dot{\theta}$ through $\partial\mathcal{S}/\partial x^\mu=p_\mu=g_{\mu\nu}\dot{x}^\nu$, can be solved by using the separation method. One gets the following equations of motion in the radial and polar angle directions:
\begin{eqnarray}
\Sigma \dot{r} &=& \pm \sqrt{\mathcal{R}(r)}\,,
\nonumber\\
\Sigma \dot{\theta} &=& \pm \sqrt{\Theta(\theta)}\,,\label{rthetadot}
\end{eqnarray}
where
\begin{eqnarray}
\mathcal{R}(r) &=& \left[ (r^2 + a^2) E - a L \right]^2 - \Delta \left[ {\cal K} + (L - a E)^2 \right]\,,
\nonumber\\
\Theta(\theta) &=& {\cal K} + \left( a^2 E^2 - \frac{L^2}{\sin^2\theta} \right) \cos^2\theta\,,\label{RTHETASEP}
\end{eqnarray}
with ${\cal K}$ being the Carter separation constant, and one can interpret $\mathcal{R}(r)$ as the effective potential in the radial motion.

For convenience, we rescale the constants of motion for photon orbits:
\bea 
\xi \equiv \frac{L}{E}\,,\qquad \eta \equiv \frac{\cal K}{E^2}\,.
\eea
The conditions for the spherical orbits, with radii collectively denoted as $r_p$, are given by
\bea
\mathcal{R}(r_p) = 0\,, \qquad \frac{d \mathcal{R}(r)}{dr}\Big|_{r_p}= 0\,.
\label{critical}
\eea
The solution to these equations is
\begin{eqnarray}
\xi(r_p)
&=& \frac{1}{a} \left( r^2 + a^2 -\frac{4 r \Delta(r)}{\Delta'(r)} \right) \Bigg|_{r=r_p}
\,, \label{xirp}\\
\eta(r_p)
&=& \frac{16 r^2 \Delta(r)}{[\Delta'(r)]^2}\Bigg|_{r=r_p}
- \left[\xi(r_p) - a\right]^2 \,.\label{etarp}
\end{eqnarray}
We thus find that the constants of motion $\xi$ and $\eta$ on a spherical photon orbit can be labeled by the radius $r_p$ of the orbit using Eqs.~\eqref{xirp} and~\eqref{etarp}. The orbits that contribute to the critical curve on the image plane should be unstable, i.e., $d^2\mathcal{R}(r)/dr^2|_{r_p}>0$.
By using Eqs.~\eqref{xirp} and \eqref{etarp}, we obtain the identity
\begin{equation}
\frac{d^2\mathcal{R}(r)}{dr^2}\Bigg|_{r_p} = -4 a r_p \frac{d\xi(r_p)}{dr_p}\,.
\label{rpplp}
\end{equation}
The values of $r_p$ for which $a\, d\xi(r_p)/dr_p < 0$ correspond to unstable spherical photon orbits, which are responsible for the formation of the shadow image.

To visualize the shadow critical curve, we define the celestial coordinates $(\alpha,\beta)$ on the image plane. The coordinate $\alpha$ is the apparent
perpendicular distance between the critical curve and the axis of symmetry, i.e., the spin axis. The coordinate $\beta$, on the other hand, is defined by the apparent perpendicular distance between the critical curve and the
plane normal to the spin axis, with the axis piercing through the plane at the black hole. We denote the position of the observer by $(r_0,\theta_0)$ with $\theta_0$ being the inclination angle between the spin axis of the black hole and the direction to the observer. For an observer in the far exterior and asymptotically flat region $r_0\rightarrow\infty$, the celestial coordinates $(\alpha,\beta)$ can be expressed as \cite{Vazquez:2003zm}
\begin{align}
\alpha&=\lim_{r_0\rightarrow\infty}\left(-r_0^2\sin\theta_0\frac{d\psi}{dr}\Bigg|_{r_0,\theta_0}\right)\,,\nonumber\\ \beta&=\lim_{r_0\rightarrow\infty}\left(r_0^2\frac{d\theta}{dr}\Bigg|_{r_0,\theta_0}\right)\,.\label{alphabetadef}
\end{align}
Using the geodesic equations Eqs.~\eqref{tphidot} and \eqref{rthetadot}, one can rewrite Eq.~\eqref{alphabetadef} in terms of the constants of motion $(\xi,\eta)$ and the inclination angle $\theta_0$ in the following expressions:\footnote{The expressions \eqref{alphabeta} can also be obtained using the tetrad formalism \cite{1973ApJ...183..237C}, by which one can define a generalized version of the celestial coordinates for observers at finite radius coordinates, see, e.g., Ref.~\cite{Frost:2024pxa}.} 
\begin{equation}
\alpha = - \frac{\xi}{\sin\theta_0}\,,\qquad \beta = \pm \sqrt{\eta + a^2 \cos^2\theta_0 -\xi^2 \cot^2\theta_0}\,.\label{alphabeta}
\end{equation}

To summarize, the procedure of finding the shadow critical curve is as follows. The celestial coordinates $(\alpha,\beta)$ are given in terms of the radii of spherical photon orbits $r_p$ through Eqs.~\eqref{xirp}, \eqref{etarp} and \eqref{alphabeta}. When the black hole is spinning, there are several spherical photon orbits, each with a different $r_p$. The orbits with different radii contribute to different points $(\alpha,\beta)$ on the image plane. We then identify the range of $r_p$ for which $a\, d\xi(r_p)/dr_p < 0$ is satisfied (see Eq.~\eqref{rpplp}). This determines the whole allowed $(\alpha, \beta)$ through Eq.~\eqref{alphabeta}, which form the shadow image of the unstable spherical photon orbits. When $a=0$, Eqs.~\eqref{xirp} and \eqref{etarp} are no longer valid. We have to use Eq.~\eqref{critical} to find the spherical photon orbits. In this case, the collection of spherical photon orbits reduces to circular orbits with different tilt angles.
These circular orbits have the same radius and share the same value of $\eta+\xi^2$.
The celestial coordinates $(\alpha,\beta)$ given in Eq.~\eqref{alphabeta} satisfy $\alpha^2+\beta^2=\eta+\xi^2$, and the critical curve on the image plane appears as a circle, independent of the inclination angle $\theta_0$.

\subsection{Shadow observables}\label{subsec:obs}

In typical scenarios, the shadow critical curves of black holes appear as a closed contour on the image plane. Its apparent size depends on the parameters of the black hole. In addition, it may not appear as a perfect circle. For instance, the Kerr black hole casts a D-shaped critical curve because of the frame-dragging effects induced by its spin. Therefore, by properly defining the shadow observables that quantify these characteristics, i.e., the apparent size and shape, one can compare the models with the real observations and place constraints on the parameter space of the models.

We first define the fractional diameter deviation $\delta$ as the relative difference between the analytic diameter of the shadow and the diameter of the Schwarzschild shadow \cite{EventHorizonTelescope:2022xqj}:
\begin{equation}
\delta\equiv\frac{\bar{d}}{6\sqrt{3}}-1\,,
\end{equation}
where $\bar{d}$ is the areal median shadow diameter defined by
\begin{equation}
\bar{d} \equiv \frac{2}{G_0 M} \sqrt{\int \frac{2\beta(r_p)}{\pi} d\alpha(r_p)}\,,
\end{equation}
where one integrates over the entire $\alpha$ axis on the image. One can recast this integration in terms of $r_p$. This integral then encompasses the entire range of unstable spherical photon orbits that contribute to the critical curve.
 
The fractional diameter deviation $\delta$ has been constrained by combining the EHT observations of the Sgr A*, as well as the accurate measurement of the mass-to-distance ratio of the black hole from the stellar motion around it. At the 68th percentile credible levels, we have \cite{EventHorizonTelescope:2022xqj}
\begin{equation}
\delta=-0.04^{+0.09}_{-0.10}\,\,\,(\textrm{Keck})\,,\qquad\delta=-0.08^{+0.09}_{-0.09}\,\,\,(\textrm{VLTI})\,,\label{deltabound}
\end{equation}
which are based on the Keck and VLTI priors on the mass-to-distance ratio, respectively.

Another quantity is the circularity deviation $\Delta C$, which is widely used in the literature to quantify the amount of distortion of the shadow critical curve. It is defined as \cite{Johannsen:2010ru,Johannsen:2013vgc}
\begin{equation}
\Delta C\equiv 2\sqrt{\frac{1}{2\pi}\int_0^{2\pi}\left(R(\phi)-\bar{R}\right)^2d\phi}\,,
\end{equation}
where 
\begin{equation}
R(\phi)=\sqrt{(\alpha-\alpha_c)^2+\left(\beta-\beta_c\right)^2}\,,\qquad \tan\phi=\frac{\beta-\beta_c}{\alpha-\alpha_c}\,.
\end{equation}
The quantity $R(\phi)$ denotes the distance between a point, labeled by the angle $\phi$, on the shadow contour and a symmetric reference point inside the contour. The typical choice of the reference point is $(\alpha_c,\beta_c)=((\alpha_L+\alpha_R)/2,0)$, where $\alpha_L$ and $\alpha_R$ represent the $\alpha$-coordinates of the leftmost and the rightmost points on the contour. The quantity $\bar{R}$ is the average radius of the contour over the angle $\phi$, i.e., $\bar{R}=\int_0^{2\pi}{R(\phi)d\phi}/2\pi$. The circularity deviation $\Delta C$ quantifies how much the shadow contour deviates from a perfect circle. If the contour is a perfect circle, the circularity deviation $\Delta C$ vanishes.

\section{Results} \label{sec:results}

After introducing the derivation of shadow critical curves and the associated observables, we will discuss the shadows of the black holes that satisfy the consistency of black hole thermodynamics in this section. In particular, we will focus on the black hole metric \eqref{ksol} with the running Newton coupling $G(r,M,a)$ parametrized by $G(z)$ with $z$ given in Eq.~\eqref{zgeneral}. We will consider the following sets of models: $(l,p)=(1,-1)$, $(1,0)$, $(1,4)$, and $(0,1)$. The main goal is twofold: one is to place constraints on the models using the observations of black hole images, and the other is to look for possible features in the shadow contours that are shared among these models. 

\subsection{Corrections without mass suppression $(l,p)=(1,-1)$}

We first consider $(l,p)=(1,-1)$, from which the variable $z$ given in Eqs.~\eqref{zgeneral} and \eqref{zgeneralrescale} becomes
\begin{equation}
z\equiv \frac{G_0Mr}{r^2+a^2}=\frac{r_*}{r_*^2+a_*^2}\,.
\label{z1}
\end{equation}
The expression of $z$ in terms of the dimensionless $r_*$ and $a_*$ does not depend on the black hole mass $M$. Therefore, the corrections induced by the running of $G(z)$ or, more specifically, the coupling parameters that control these corrections, can be directly constrained through black hole observations, irrespective of the mass of the black holes.

For generality, we assume $G(z)$ to be of the form of Pad\'e expansion with respect to $z$
\begin{equation}
G(z)\equiv G_0\left(\frac{1+\sum_{n=1}a_nz^n}{1+\sum_{m=1}b_mz^m}\right)\,,
\label{gexp1m1}
\end{equation}
where the indices $(m,n)\in\mathbb{Z}$. In this parametrization \eqref{gexp1m1}, we have a set of coupling constants $(a_i,b_i)$ which do not depend on $M$ and are truly dimensionless. As we mentioned, the constraining power on $(a_i,b_i)$ will not be suppressed by the black hole mass $M$ when considering supermassive black hole images. 

We would like to mention that because for Eq.~\eqref{z1}, we have $|z|\rightarrow0$ when $|r_*|\rightarrow0$ or $|r_*|\rightarrow\infty$, the asymptotic flatness condition ($G(z)\rightarrow G_0$ when $|r_*|\rightarrow\infty$) and the regularity condition at the origin ($G(z)\rightarrow 0$ when $|r_*|\rightarrow0$) cannot be satisfied simultaneously. Therefore, the asymptotic flatness condition at spatial infinity implies that the spacetime property near the origin, including the singular behavior, retains its Kerr nature. In addition, we confirm that, with the ansatz \eqref{gexp1m1}, the ADM mass of the spacetime is still given by $M$. The asymptotic expansion of $g_{tt}$ component reads
\begin{equation}
g_{tt}=-1+\frac{2}{r_*}+\frac{2\left(a_1-b_1\right)}{r_*^2}+\mathcal{O}\left(r_*^{-3}\right)\,.
\end{equation}
Therefore, to satisfy the Solar System constraint \cite{Williams:2004qba}, it is suitable to assume $a_1=b_1$. This then requires higher-order coefficients $a_i\ne b_i$ in order to go beyond Kerr in this model. Considering expansion up to $z^2$, we can rewrite Eq.~\eqref{gexp1m1} as
\begin{equation}
G(z) = G_0 \left( \frac{1 + a_2 z^2/(1 + b_1 z)}{1 + b_2 z^2/(1 + b_1 z)}\right)\,,\label{Gm1ex}
\end{equation}
from which it can be seen that as $b_1\gg1$, the Newton coupling $G(z)$ reduces to $G_0$, irrespective of the values of $a_2$ and $b_2$.

\begin{figure*}[htb]
\centering
\includegraphics[width=230pt]{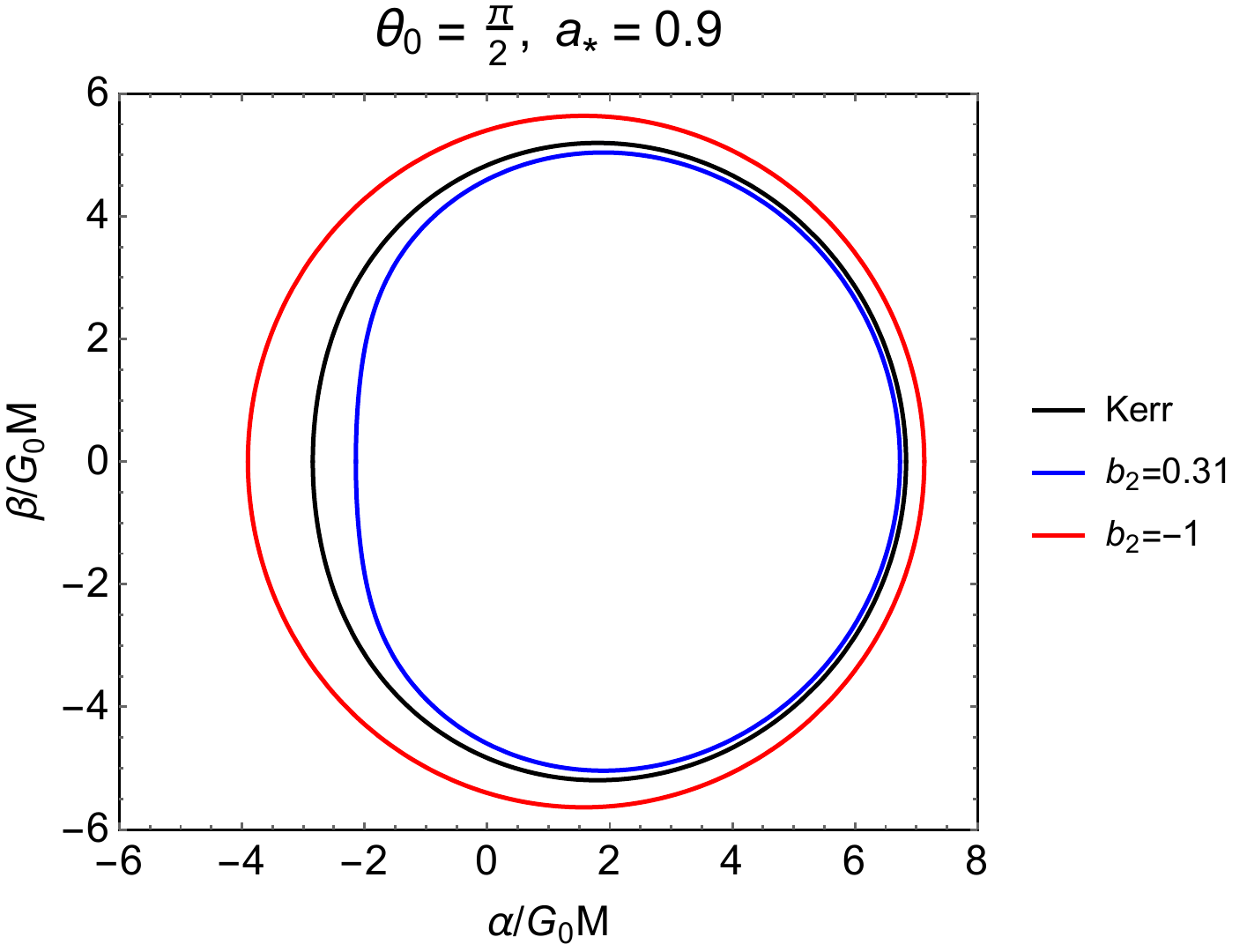}
\includegraphics[width=230pt]{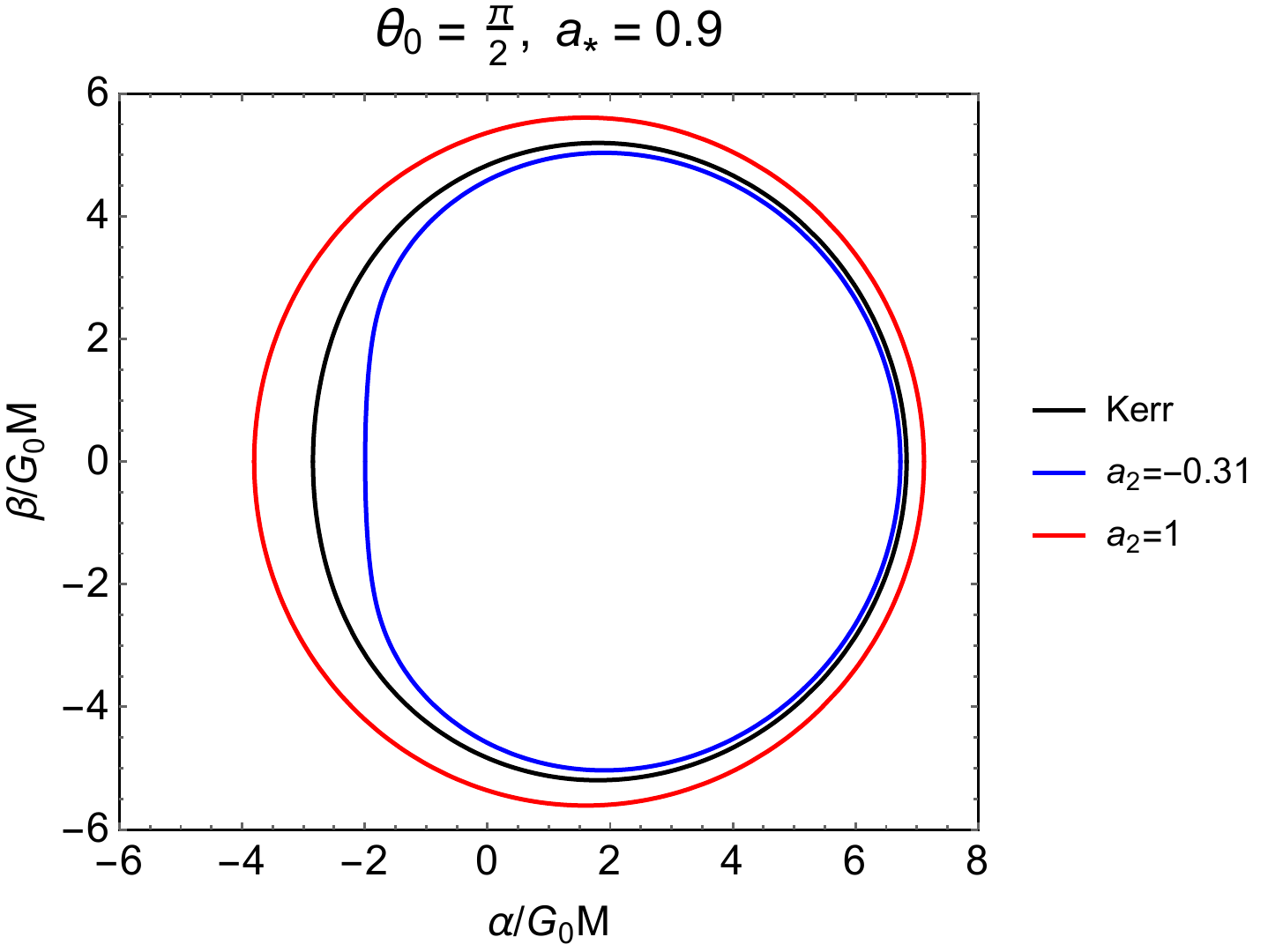}
\caption{The shadow critical curves (left) for $b_1=a_2=0$ and different values of $b_2$ and (right) for $b_1=b_2=0$ and different values of $a_2$.}
\label{fig:criticalcurvedm11}
\end{figure*}

\begin{figure*}[htb]
\centering
\includegraphics[width=230pt]{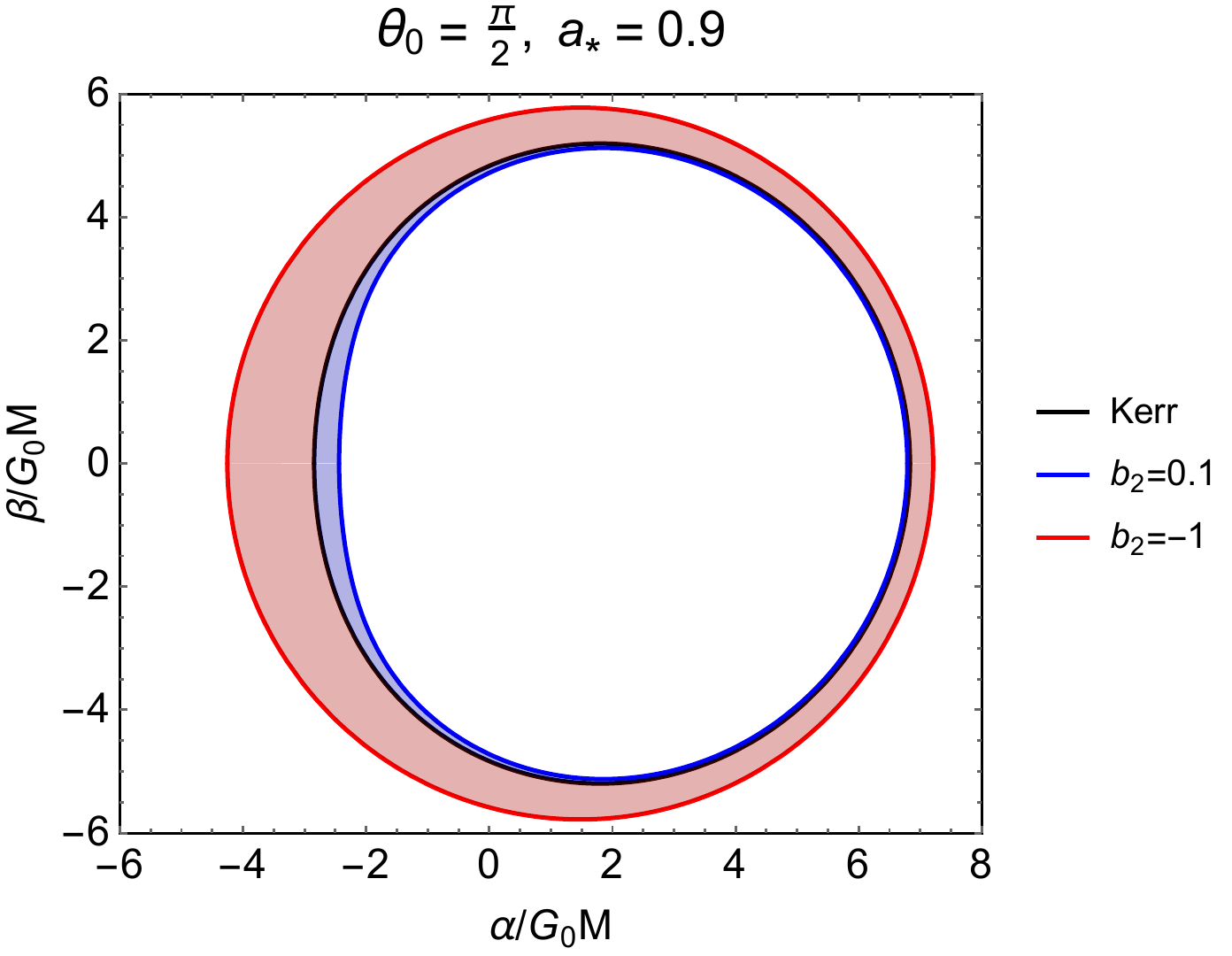}
\includegraphics[width=230pt]{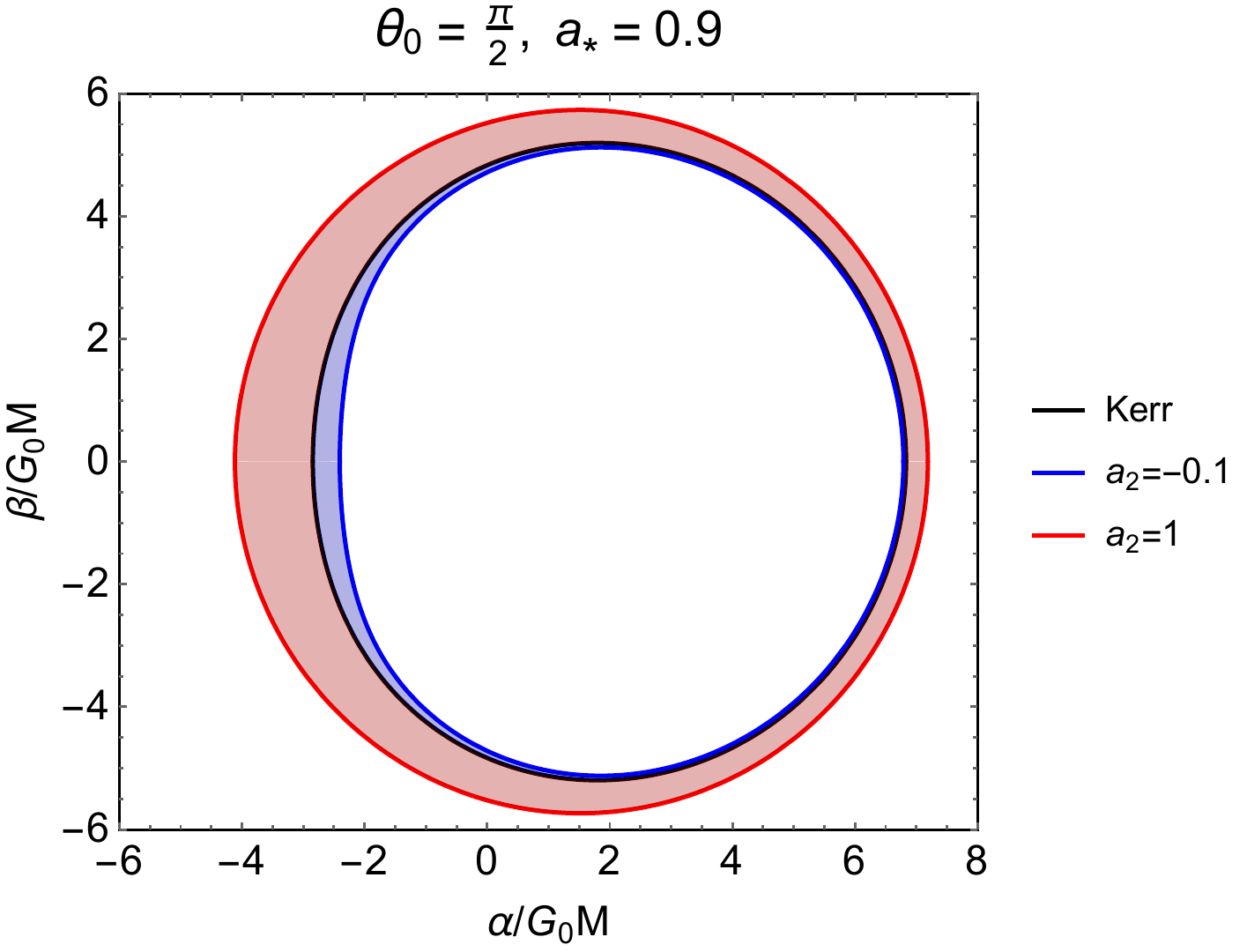}
\caption{Left: the critical curves for fixed $a_2=0$;
the blue and the red curves with $(b_1,b_2)=(-1,0.1)$ and $(b_1,b_2)=(-1,-1)$, respectively. Right: the critical curves for fixed $b_2=0$; the blue and the red curves with $(b_1,a_2)=(-1,-0.1)$ and $(b_1,a_2)=(-1,1)$, respectively. In each panel, the shaded regions represent the results of contours scanned when taking $b_1\ge-1$. For $b_1\rightarrow\infty$, the critical curves converge to the Kerr results (black curves).}
\label{fig:criticalcurved2}
\end{figure*}

From now on, we will fix $\theta_0=\pi/2$. Choosing $a_*=0.9$ and $b_1=0$, in Fig.~\ref{fig:criticalcurvedm11}, we first show the shadow critical curves for different values of $(a_2,b_2)$. On the left panel, we fix $a_2=0$ and vary $b_2$, while on the right panel we fix $b_2$ and vary $a_2$. The critical curve of the Kerr shadow is shown by the black contour. One can see that the black hole critical curves in the models with $(a_2,b_2)=(0,k)$ are very similar to those of $(a_2,b_2)=(-k,0)$. Such a degeneracy can be understood by taking the Taylor expansion of the right-hand side of Eq.~\eqref{Gm1ex} near $z=0$, which gives $G(z)/G_0\approx 1+(a_2-b_2)z^2$. In addition, increasing the value of $b_1$ suppresses the corrections induced by $(a_2,b_2)$. In the limit of $b_1\rightarrow\infty$, we have $G(z)\rightarrow G_0$ and the Kerr spacetime is recovered, irrespective of the values of $(a_2,b_2)$ (Fig.~\ref{fig:criticalcurved2}).

\begin{figure*}[htb]
\centering
\includegraphics[width=230pt]{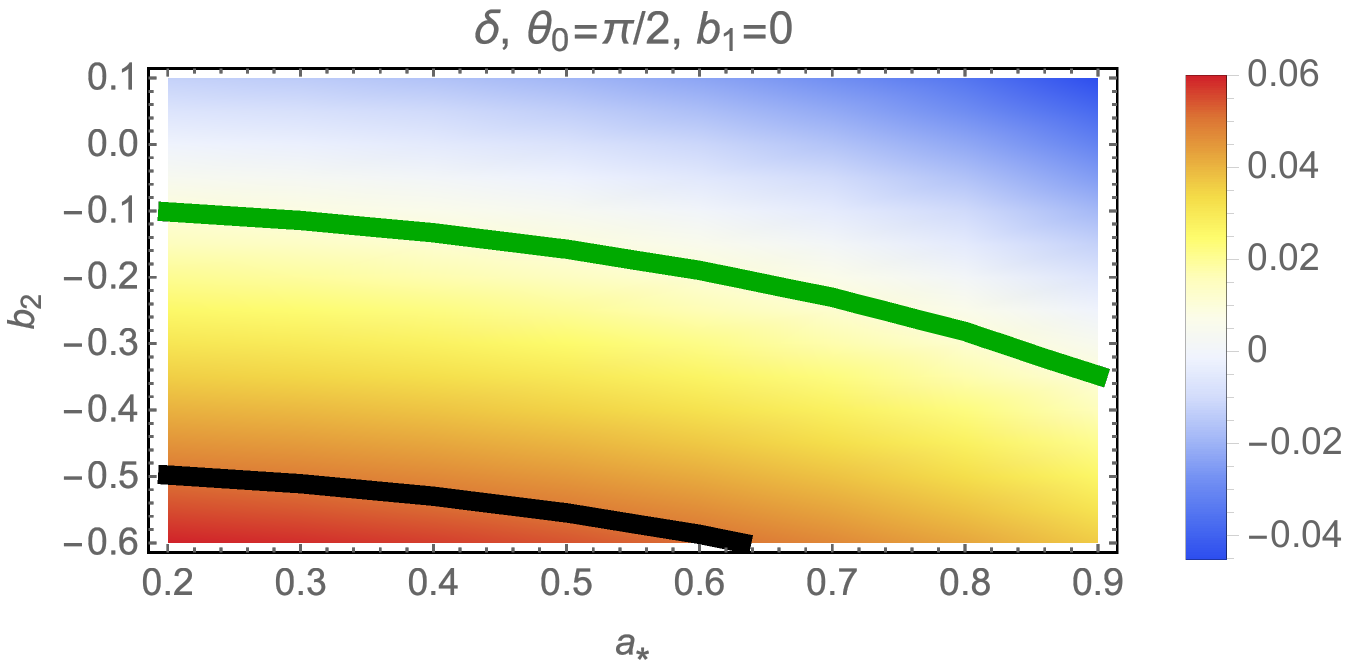}
\includegraphics[width=230pt]{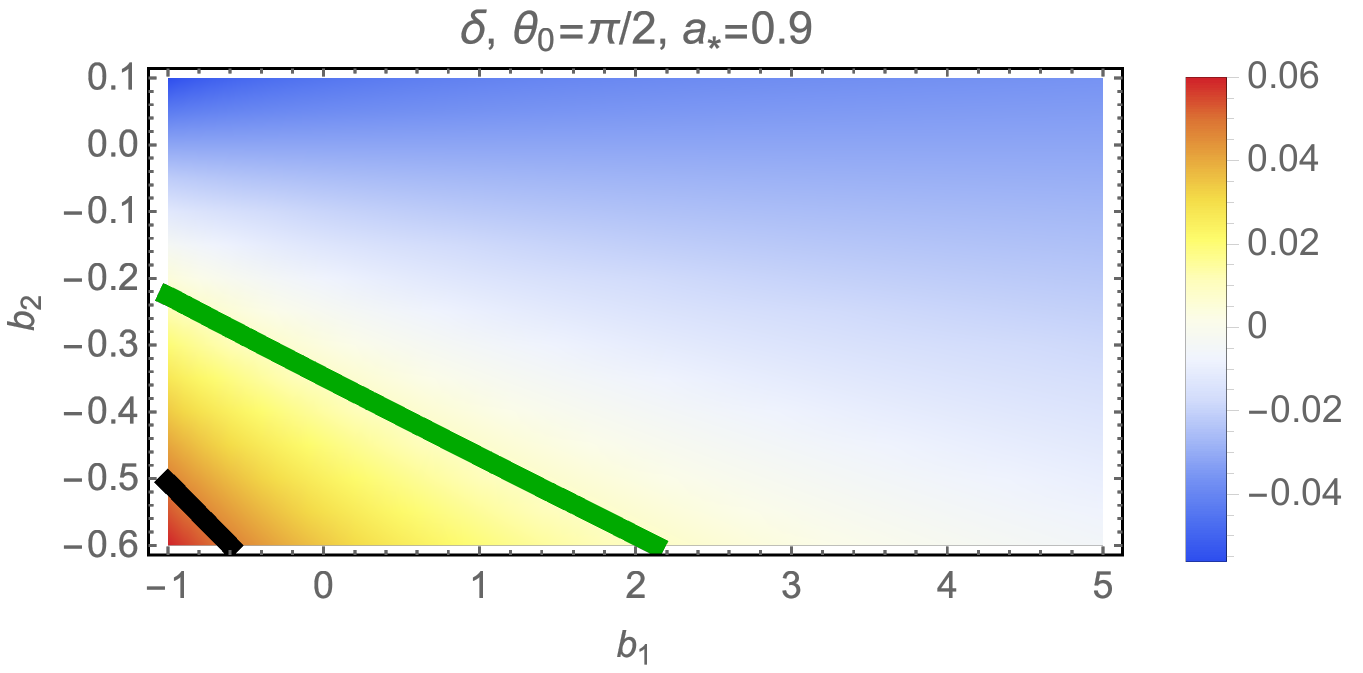}
\caption{The fractional diameter deviation $\delta$ is shown by color in the parameter space of $(a_*,b_2)$ (left) and $(b_1,b_2)$ (right). The black and green curves represent the upper bounds obtained from the Keck and VLTI priors on the mass-to-distance ratio of the Sgr A*, i.e., Eq.~\eqref{deltabound}. The parameter space toward the blue region is more observationally preferred.}
\label{fig:Rfdensitym1}
\end{figure*}

\begin{figure*}[htb]
\centering
\includegraphics[width=230pt]{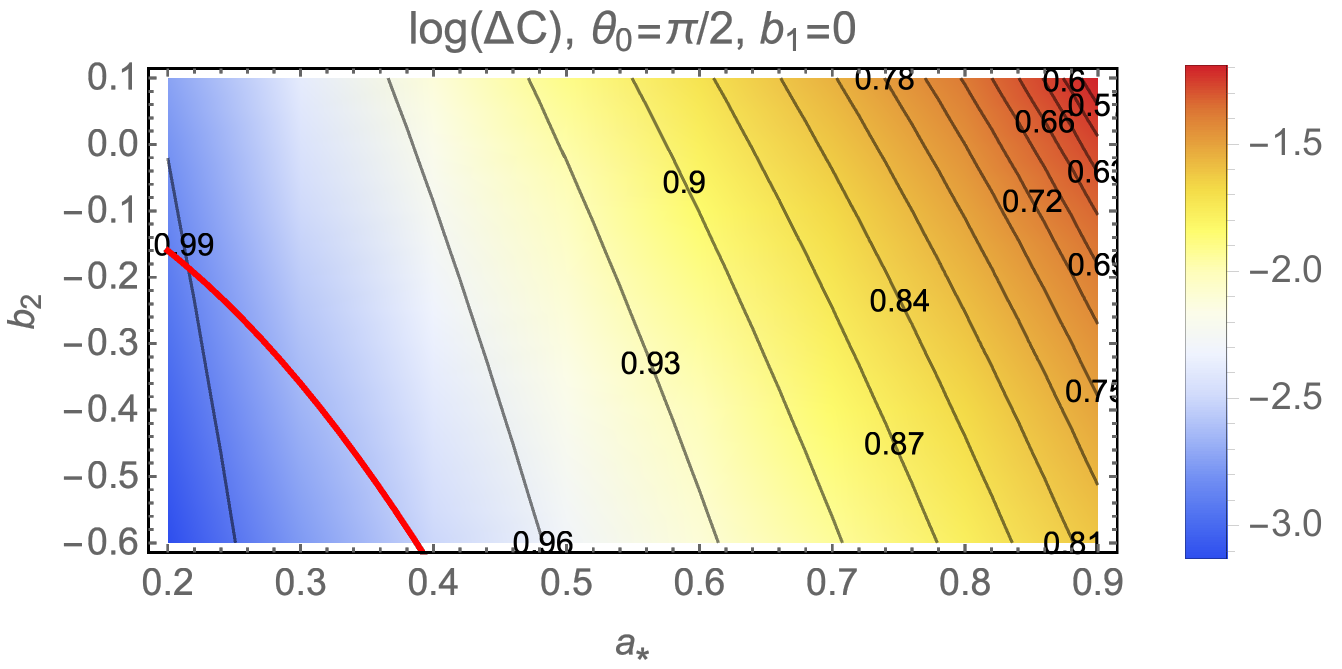}
\includegraphics[width=230pt]{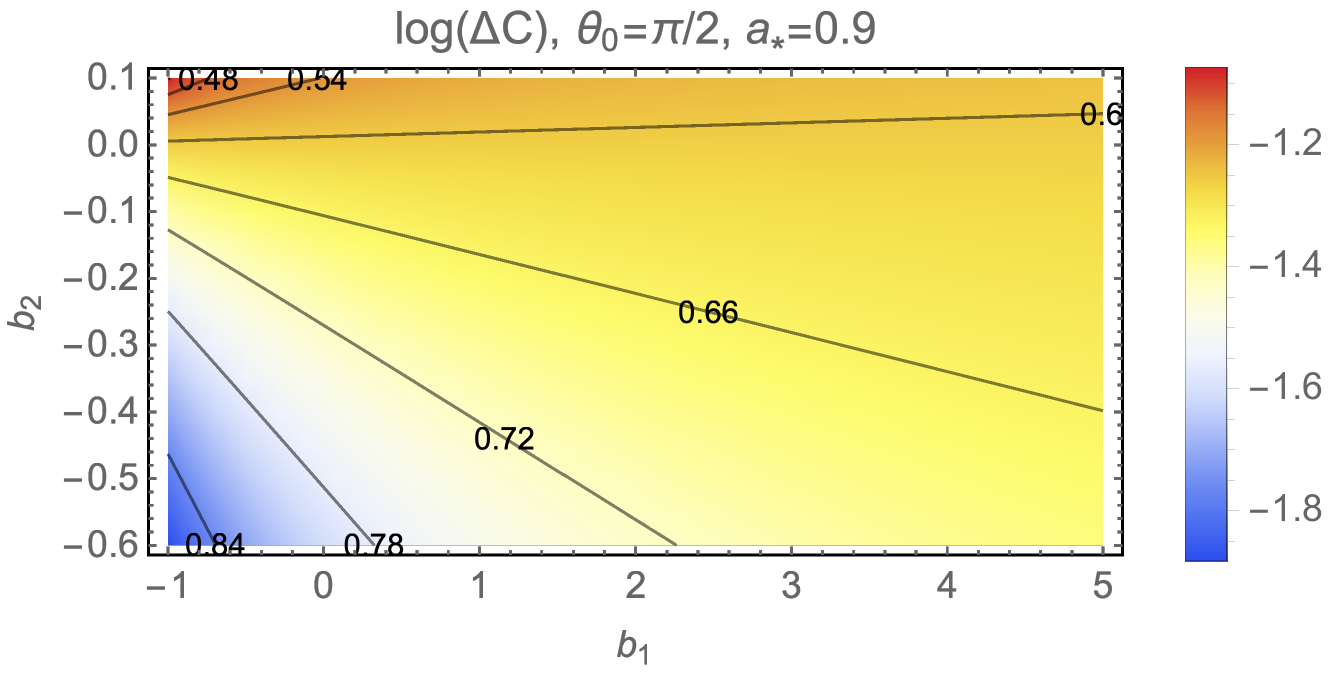}
\caption{The logarithm of the circularity deviation $\Delta C$  is shown by color in the parameter space of $(a_*,b_2)$ (left) and $(b_1,b_2)$ (right). The black contours represent the values of the deviation parameter $\epsilon$ from extremality defined in Eq.~\eqref{defepsilon}. On the left panel, the region to the left of the red curve represents the black hole spacetimes which only have one event horizon.}
\label{fig:dCdensitym1}
\end{figure*}

As one can see from Eq.~\eqref{gexp1m1} that the choice of the coefficients $(a_i,b_i)$ may increase or decrease the Newton coupling $G(z)$, which can directly alter the size of the critical curves. In Fig.~\ref{fig:Rfdensitym1}, we show the fractional diameter deviation $\delta$ in the parameter space $(a_*,b_2)$ and $(b_1,b_2)$ on the left and right panels, respectively, for $a_1=b_1$ and $a_2=0$. On the left panel, we choose $a_1=b_1=0$, while on the right panel, we fix $a_*=0.9$. The black and green curves indicate the constraints on $\delta$ from the upper limit of the Keck and VLTI measurements, respectively. The parameter space toward the blue region is more preferred observationally.

In Fig.~\ref{fig:dCdensitym1}, we show the circularity deviation $\Delta C$ (in logarithmic scale) in the parameter space of $(a_*,b_2)$ (left) and $(b_1,b_2)$ (right) for fixed $a_2=0$. Similar to Fig.~\ref{fig:Rfdensitym1}, we fix $a_1=b_1=0$ on the left panel, and $a_*=0.9$ on the right panel, respectively. The black curves are the contours of the extremality deviation $\epsilon$ defined in Eq.~\eqref{defepsilon}. With the results shown in Fig.~\ref{fig:Rfdensitym1}, we find that shrinking the shadow size, which can be achieved by decreasing the Newton coupling (e.g., the blue contours in Fig.~\ref{fig:criticalcurvedm11} where $b_2>0$ or $a_2<0$) or by increasing $a_*$, would lead to a more distorted shadow contour, resulting in the increase of $\Delta C$. In addition, we find that the contours of circularity deviation $\Delta C$ follow perfectly with those of extremality deviation $\epsilon$ for the case of $(l,p)=(1,-1)$ in the parameter space we consider in Fig.~\ref{fig:dCdensitym1}. In particular, when some coefficients $b_i$ are negative, the function $\Delta(r)$ may have a pole between the zeros such that the spacetime only has one event horizon $r_h$. We exhibit such regions of parameter space on the left panel of Fig.~\ref{fig:dCdensitym1}, i.e., the region to the left of the red curve. It turns out that even in this region, the circularity deviation $\Delta C$ of the shadow still matches well the extremality deviation $\epsilon$.

\subsection{Mass-independent scale identification $(l,p)=(1,0)$}

Here, we consider the model with $(l,p)=(1,0)$, which belongs to the class of mass-independent scale identification in the FRG method. In this case, Eqs.~\eqref{zgeneral} and \eqref{zgeneralrescale} can be rewritten as
\begin{equation}
z=\frac{G_0}{r^2+a^2}=\frac{1}{G_0M^2\left(r_*^2+a_*^2\right)}\,.
\end{equation}
Again, for generality, we assume $G(z)$ to be of the form of Pad\'e expansion with respect to $z$,
\begin{equation}
G(z)\equiv G_0\left(\frac{1+\sum_{n=1}c_nz^n}{1+\sum_{m=1}d_mz^m}\right)\,,
\label{gexp}
\end{equation}
where the indices $(m,n)\in\mathbb{Z}$ and we have a corresponding set of coupling constants $(c_i,d_i)$.{\footnote{One simple mass-independent scale identification discussed in Ref.~\cite{Chen:2022xjk} corresponds to $c_n=0$ for all $n$, $d_1=\tilde\omega$, and $d_m=0$ for $m\ne1$.}} The ansatz~\eqref{gexp} for $G(z)$ ensures that it recovers Newton constant $G_0$ when $|r|\rightarrow\infty$. In fact, when incorporating Eq.~\eqref{gexp} into the metric~\eqref{ksol}, the ADM mass is still given by $M$, and the spacetime satisfies the Solar System constraints~\cite{Williams:2004qba}. Also, the coupling $G(z)$ is symmetric with respect to $r=0$, although the ring on the surface $r = 0$ is singular.

\begin{figure*}[htb]
\centering
\includegraphics[width=230pt]{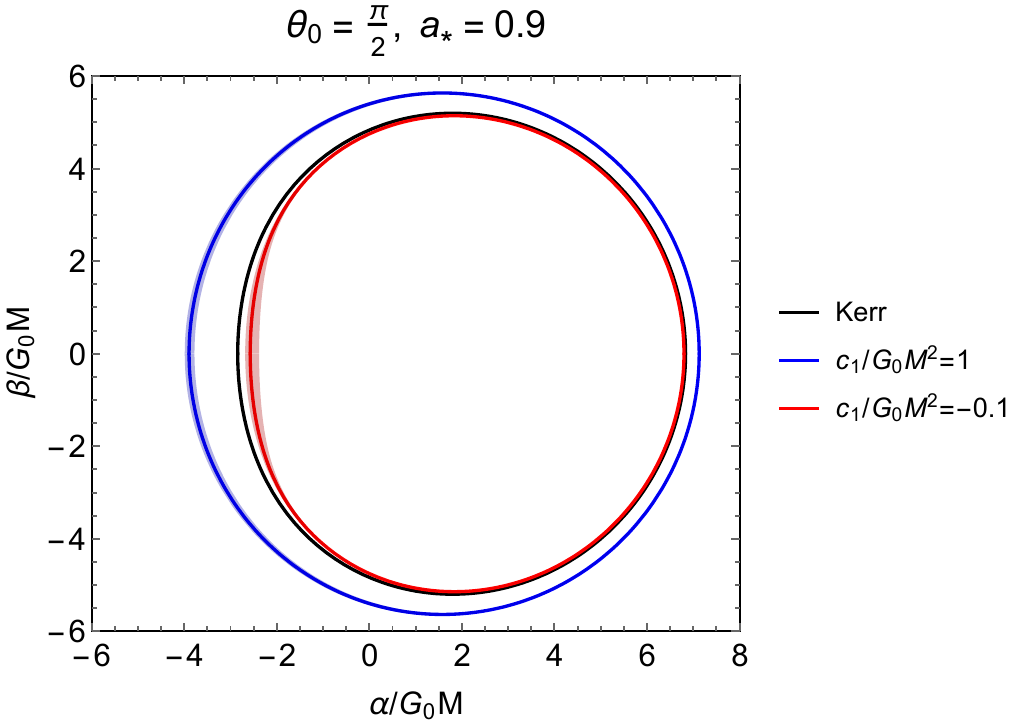}
\includegraphics[width=230pt]{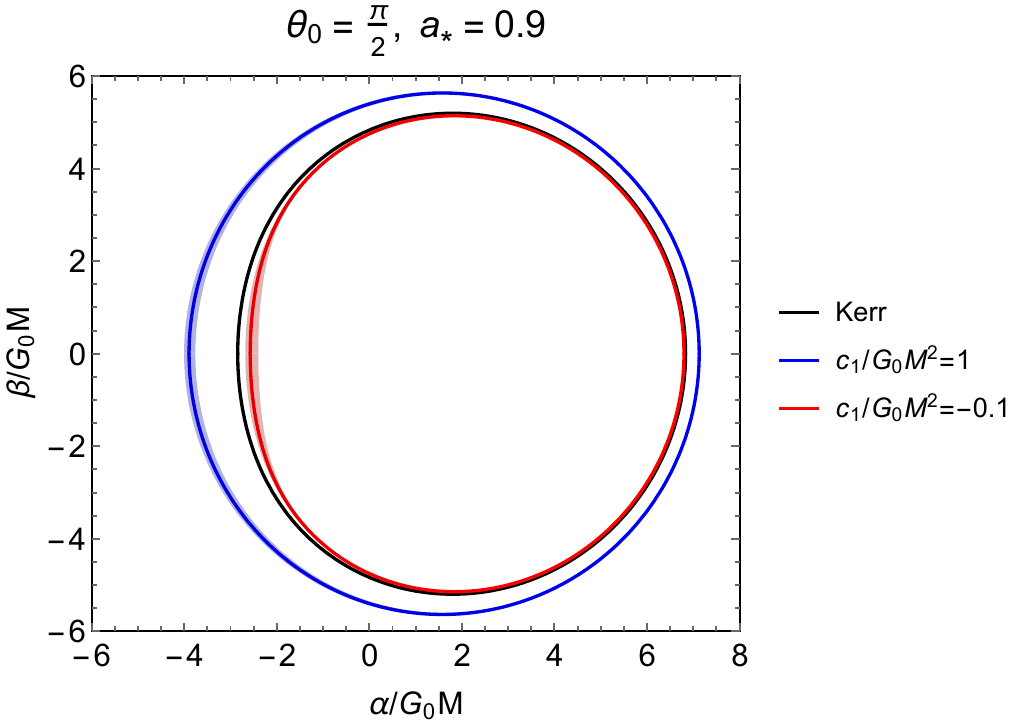}
\caption{The critical curves with $d_1=0$ and different choices of $(c_1/G_0M^2,c_2/G_0^2M^4,d_2/G_0^2M^4)$. The shaded regions on the left and right panels correspond to the results of contours scanned for varied $c_2/G_0^2M^4$ and $d_2/G_0^2M^4$, respectively. The red-shaded region on the left (right) panel represents the results with $c_2/G_0^2M^4\in(-0.1,0.1)$ ($d_2/G_0^2M^4\in(-0.1,0.1)$). On the other hand, the blue-shaded region on the left (right) panel represents the results with $c_2/G_0^2M^4\in(-1,1)$ ($d_2/G_0^2M^4\in(-1,1)$).}
\label{fig:criticalcurvec1}
\end{figure*}

\begin{figure*}[htb]
\centering
\includegraphics[width=230pt]{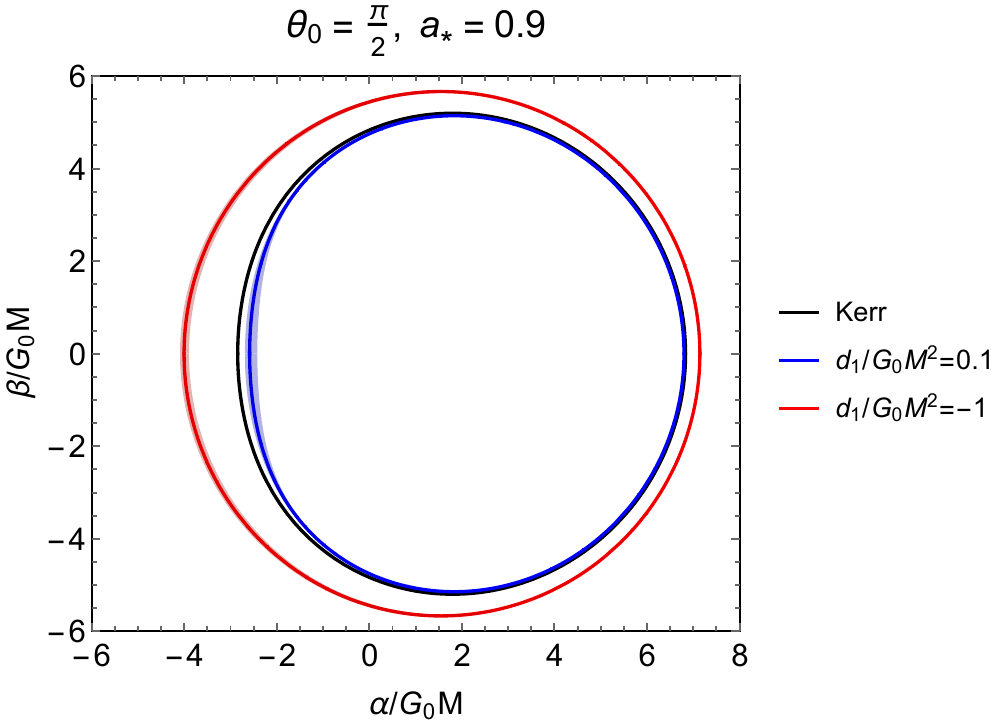}
\includegraphics[width=230pt]{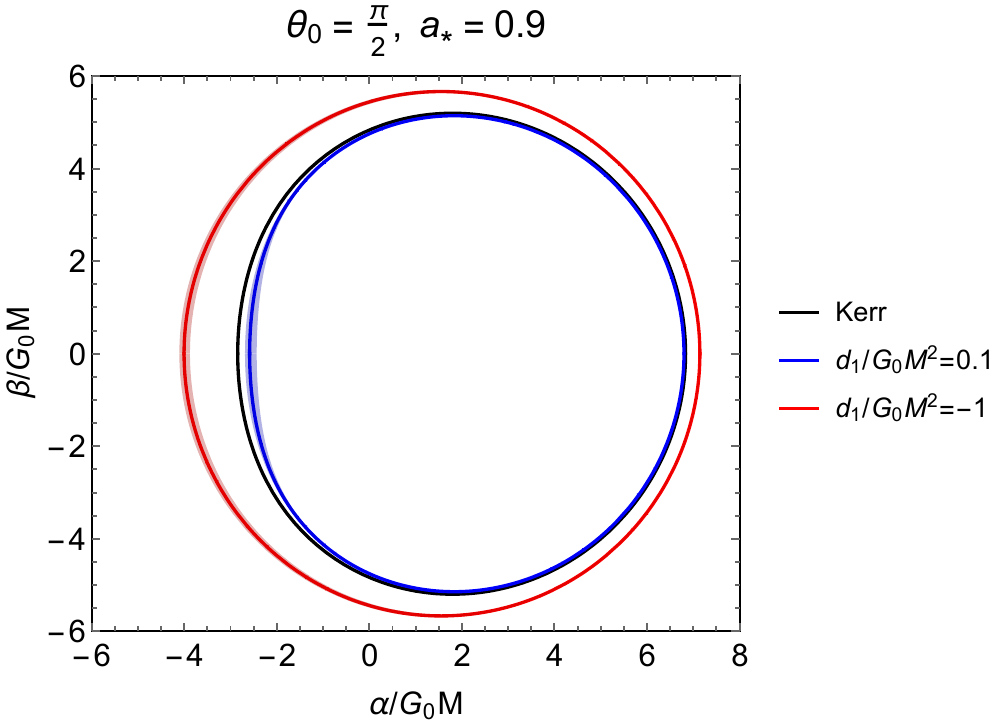}
\caption{The critical curves with $c_1=0$ and different choices of $(d_1/G_0M^2,c_2/G_0^2M^4,d_2/G_0^2M^4)$. The shaded regions on the left and right panels correspond to the results of contours scanned for varied $c_2/G_0^2M^4$ and $d_2/G_0^2M^4$, respectively. The red-shaded region on the left (right) panel represents the results with $c_2/G_0^2M^4\in(-1,1)$ ($d_2/G_0^2M^4\in(-1,1)$). On the other hand, the blue-shaded region on the left (right) panel represents the results with $c_2/G_0^2M^4\in(-0.1,0.1)$ ($d_2/G_0^2M^4\in(-0.1,0.1)$).}
\label{fig:criticalcurved1}
\end{figure*}

Assuming $a_*=0.9$ and an edge-on inclination, i.e., $\theta_0=\pi/2$, we show the critical curves with different choices of $(c_1/G_0M^2,d_1/G_0M^2,c_2/G_0^2M^4,d_2/G_0^2M^4)$ in Figs.~\ref{fig:criticalcurvec1} and \ref{fig:criticalcurved1}. We first notice that the contributions from $c_2$, $d_2$, and higher-order coupling constants are subdominant as compared with those from $c_1$ and $d_1$. One reason for this is because of the mass suppression of these higher-order corrections. Another reason is that the factor $1/(r_*^2+a_*^2)$ evaluated on the spherical photon orbits is generically smaller than unity,{\footnote{For example, it is $1/9$ for Schwarzschild black holes and $1/2$ for extreme Kerr black holes.}} such that it also suppresses the higher-order corrections. Therefore, from now on, we will focus on the corrections induced by the parameters $c_1$ and $d_1$.

In addition to the hierarchy between different orders of coefficients, one can see from Figs.~\ref{fig:criticalcurvec1} and~\ref{fig:criticalcurved1} that if the coefficients are chosen in a way such that $G(z)<G_0$, the critical curves shrink, acquire more distortion, and display a more D-shaped contour, e.g., the red contour in Fig.~\ref{fig:criticalcurvec1} where $c_1<0$ and the blue contour in Fig.~\ref{fig:criticalcurved1} where $d_1>0$. Essentially, when fixing $a_*$, the decrease of the running Newton coupling implies a smaller event horizon, and the geometry is getting closer to the extreme configuration.

\begin{figure*}[htb]
\centering
\includegraphics[width=230pt]{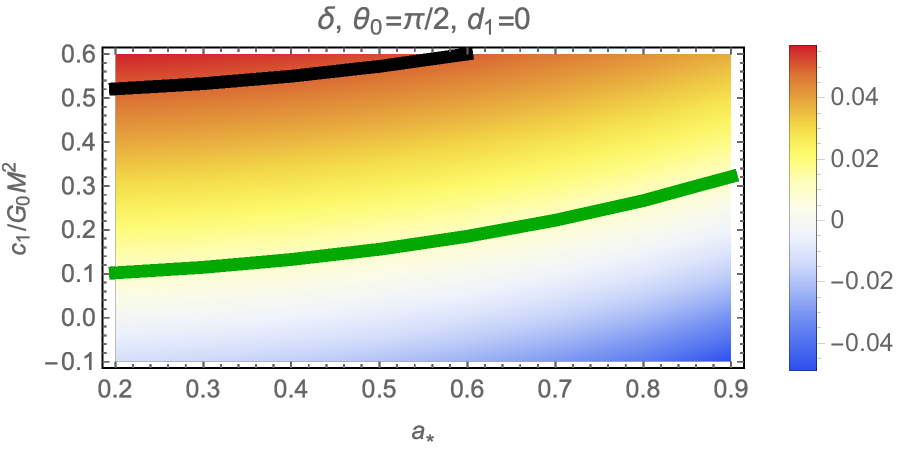}
\includegraphics[width=230pt]{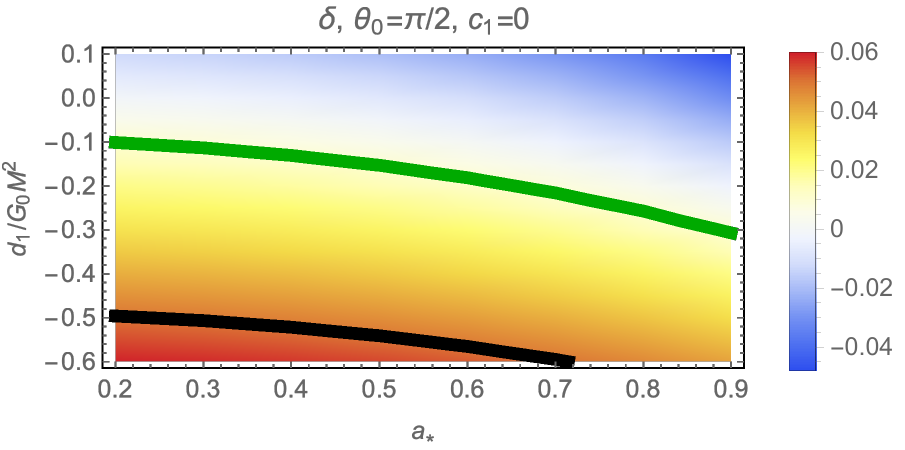}
\caption{The fractional diameter deviation $\delta$ is shown by color in the parameter space of $(a_*,c_1/G_0M^2)$ (left) and $(a_*,d_1/G_0M^2)$ (right). The upper bounds on $\delta$ using the Keck (black) and VLTI (green) priors on Sgr A* mass-to-distance ratio are presented.}
\label{fig:Rfdensitymassind}
\end{figure*}

\begin{figure*}[htb]
\centering
\includegraphics[width=230pt]{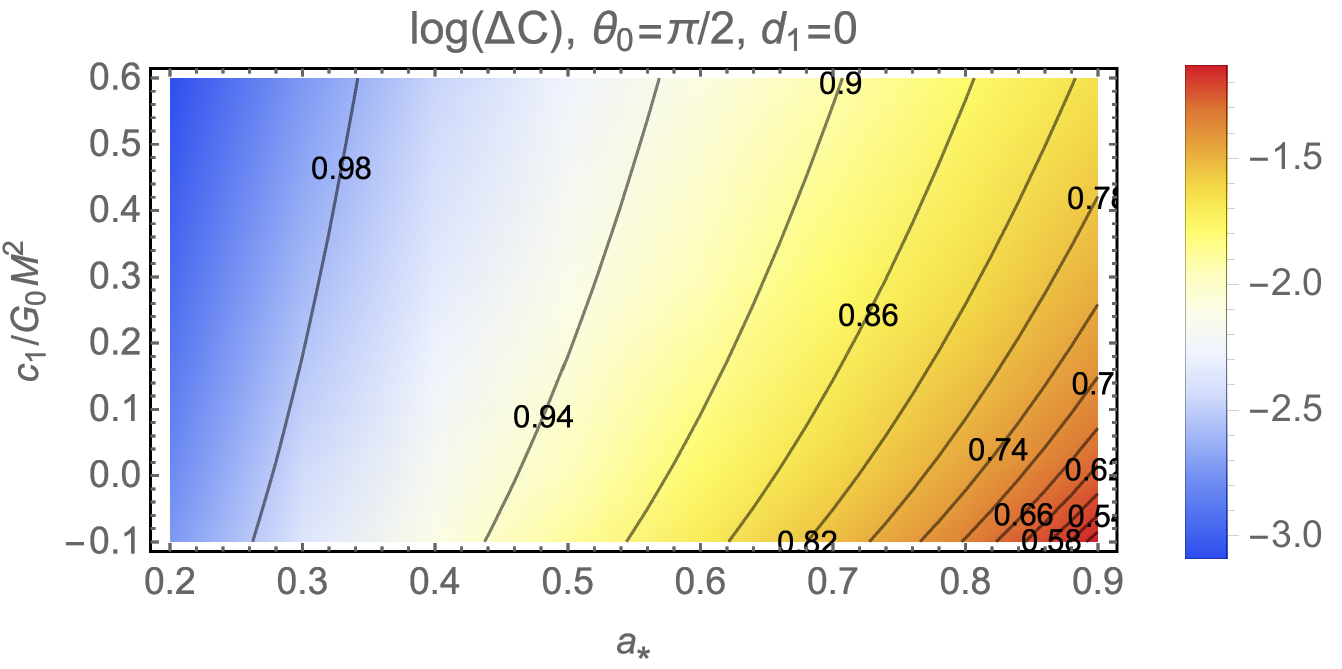}
\includegraphics[width=230pt]{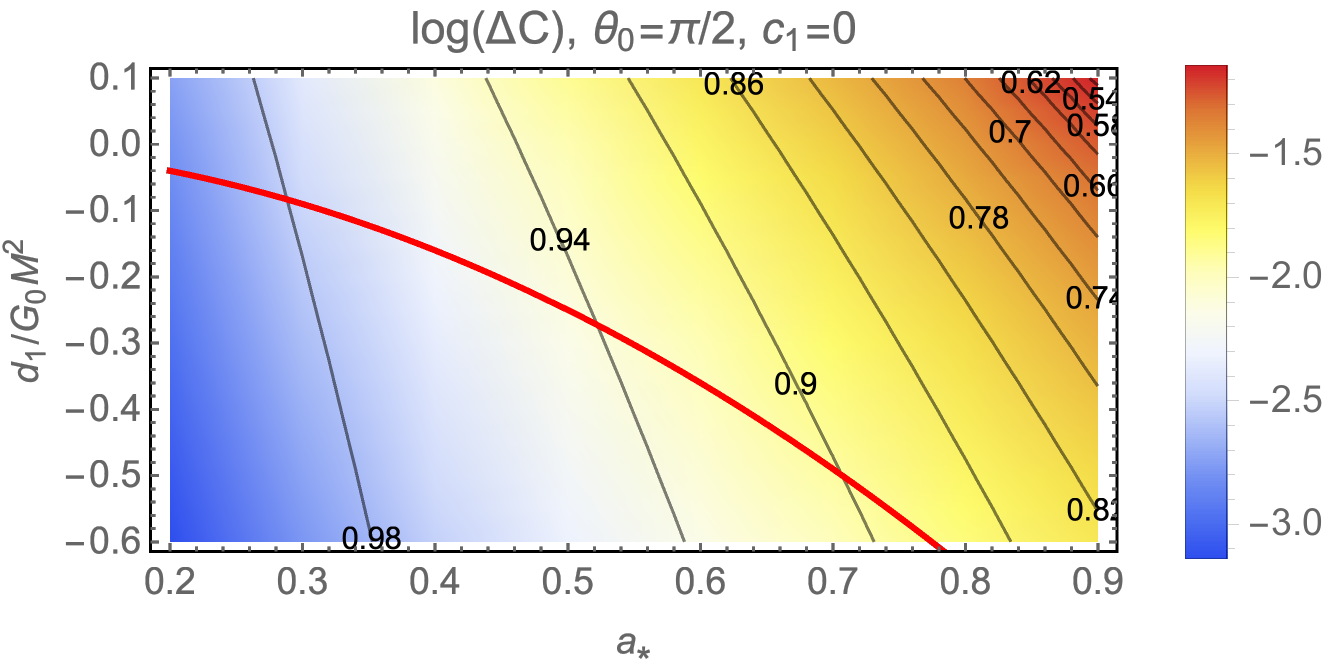}
\caption{The logarithm of the circularity deviation $\Delta C$ is shown by color in the parameter space of $(a_*,c_1/G_0M^2)$ (left) and $(a_*,d_1/G_0M^2)$ (right). The black contours represent the values of the extremality deviation $\epsilon$. On the right panel, the region to the left of the red curve represents the black hole spacetimes that only have one event horizon.}
\label{fig:dCdensitymassind}
\end{figure*}

The choice of the coefficients $(c_i,d_i)$ can directly alter the size of the critical curves. In Fig.~\ref{fig:Rfdensitymassind}, we show the fractional diameter deviation $\delta$ in the parameter space $(a_*,c_1/G_0M^2)$ and $(a_*,d_1/G_0M^2)$. The black and green curves indicate the constraints on $\delta$ from the upper limit of the Keck and VLTI measurements, respectively.

In Fig.~\ref{fig:dCdensitymassind}, we show the circularity deviation $\Delta C$ (in logarithmic scale) in the parameter space of $(a_*,c_1/G_0M^2)$ (left) and $(a_*,d_1/G_0M^2)$ (right). The black curves represent the contours of the extremality deviation $\epsilon$ defined in Eq.~\eqref{defepsilon}. Combining with the results shown in Fig.~\ref{fig:Rfdensitymassind}, we find that shrinking the shadow size, which can be achieved by decreasing the Newton coupling or by increasing $a_*$, would lead to a more distorted shadow contour and increase $\Delta C$. In addition, we find that the contours of circularity deviation $\Delta C$ follow perfectly with those of extremality deviation $\epsilon$ for the case of $(l,p)=(1,0)$ in the parameter space shown in Fig.~\ref{fig:dCdensitymassind}. This is true even in the region of the parameter space where there is only one event horizon, which could occur when some coefficients $d_i$ are negative. In this region, the function $\Delta(r)$ has a pole between the zeros. We show such a region on the right panel of Fig.~\ref{fig:dCdensitymassind}, i.e., the region to the left of the red curve.

\subsection{Models inspired by the scale identification for nonsingular black holes $(l,p)=(1,4)$}\label{subsec:p4}

Inspired by the models of nonsingular black holes proposed in Ref.~\cite{Chen:2023wdg}, we consider $(l,p)=(1,4)$ in this subsection. From Eq.~\eqref{zgeneralrescale}, it follows that
\begin{equation}
z=\frac{1}{\left(G_0M^2\right)^5\left(r_*^2+a_*^2\right)r_*^4}\,.
\end{equation}
As we have done in the previous subsections, we choose the Padé expansion for{\footnote{The nonsingular black hole model considered in Ref.~\cite{Chen:2023wdg} corresponds to a specific choice of parameters in which $e_n=0$ for all $n$, $f_1=\tilde\omega$, and $f_m=0$ for $m>1$.}} $G(z)$
\begin{equation}
G(z)\equiv G_0\left(\frac{1+\sum_{n=1}e_nz^n}{1+\sum_{m=1}f_mz^m}\right)\,.
\label{gexp2}
\end{equation}
In this case, $z$ scales roughly as $1/r_*^6$ as $r_*$ increases. The contributions from higher-order coefficients drop very quickly when one moves away from the black hole. Therefore, we neglect all the higher-order coefficients and only consider the coefficients $e_1$ and $f_1$.

\begin{figure*}[htb]
\centering
\includegraphics[width=230pt]{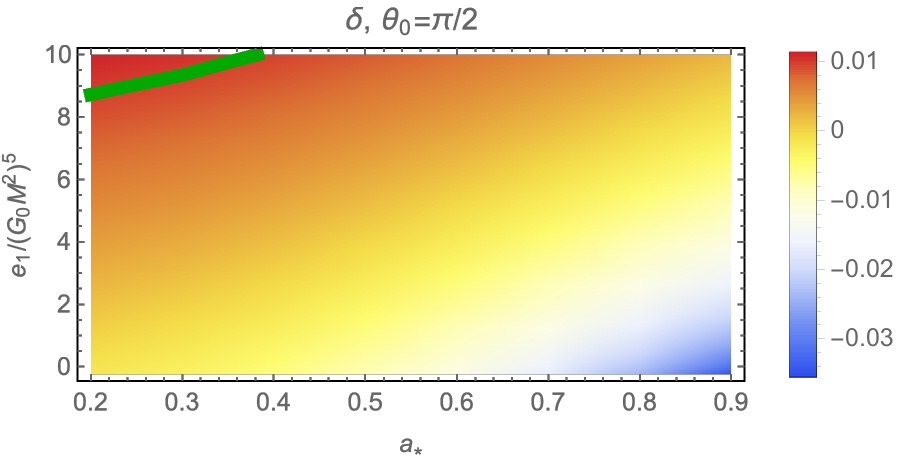}
\includegraphics[width=230pt]{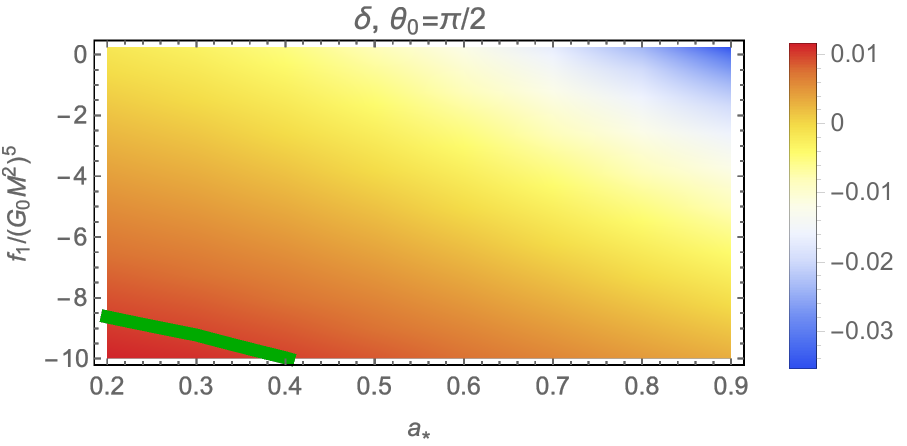}
\caption{The fractional diameter deviation $\delta$ is shown by color in the parameter space of $(a_*,e_1/(G_0M^2)^5)$ (left) and $(a_*,f_1/(G_0M^2)^5)$ (right). The green curve represents the upper bound on $\delta$ inferred from the VLTI priors.}
\label{fig:Rfdensitynonsingular}
\end{figure*}

In Fig.~\ref{fig:Rfdensitynonsingular}, we show the fractional diameter deviation $\delta$ in the parameter space of $(a_*,e_1/(G_0M^2)^5)$ and $(a_*,f_1/(G_0M^2)^5)$. The constraints from the Sgr A* shadow size with the Keck and VLTI priors on the mass-to-distance ratio are also presented. Again, decreasing $G(z)$ by either taking $e_1<0$ or $f_1>0$ reduces the size of shadows. In addition, one can see that the constraining power of the Sgr A* images on $e_1$ and $f_1$ is relatively weaker than that on the coefficients in the previous cases. This is because in the case of $(l,p)=(1,4)$, the variable $z$ decreases very quickly as $r$ increases. In addition, the corrections induced by $e_1$ and $f_1$ are strongly suppressed by the mass $M$ of the black hole. This is consistent with the expectation that the quantum effects that are responsible for resolving spacetime singularities deep inside the horizon, if they leak out of the horizon, are usually strongly suppressed by some positive powers of $l_p/r_h$, where $l_p$ is the Planck length. For supermassive black holes, such effects of Planck scales are hardly measurable outside the event horizon. 

\begin{figure*}[htb]
\centering
\includegraphics[width=230pt]{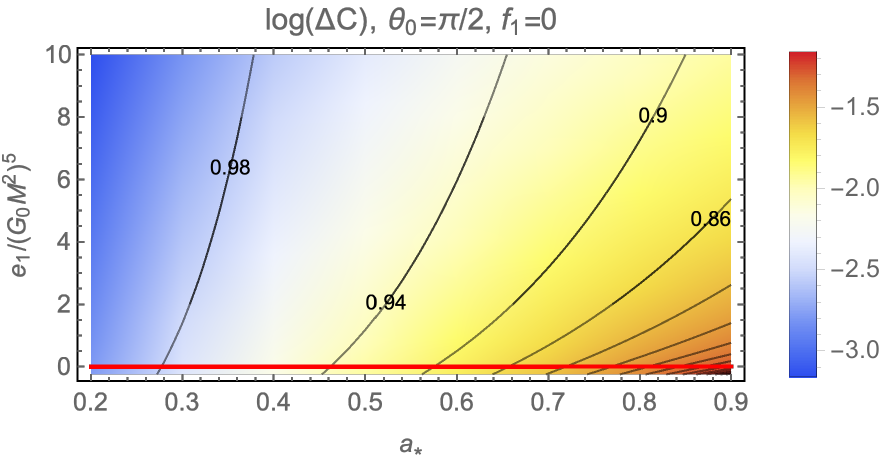}
\includegraphics[width=230pt]{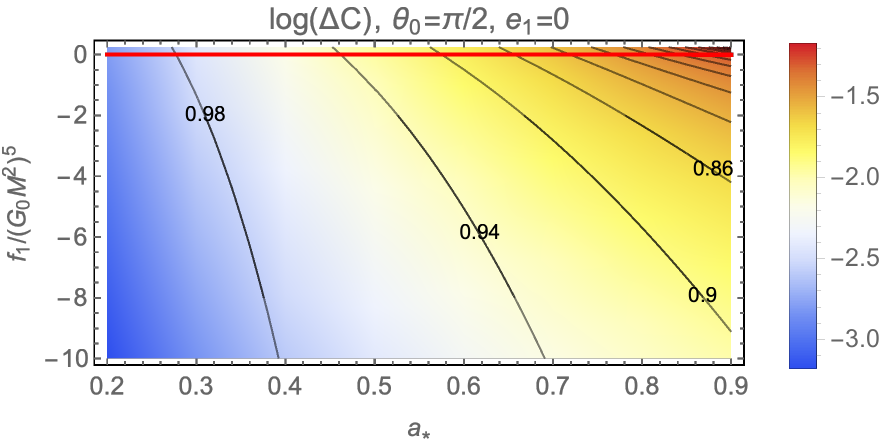}
\caption{The logarithm of the circularity deviation $\Delta C$ is shown by color in the parameter space of $(a_*,e_1/(G_0M^2)^5)$ (left) and $(a_*,f_1/(G_0M^2)^5)$ (right). The black contours represent the values of the extremality deviation $\epsilon$. The parameter space above (below) the red curve on the left (right) panel corresponds to the black hole spacetimes that have only one event horizon.}
\label{fig:dCdensitynonsingular}
\end{figure*}

We would like to emphasize that the spacetime structure in the model~\eqref{gexp2} can be quite different from the previous cases with $(l,p)=(1,-1)$ or $(l,p)=(1,0)$. For instance, when $e_1>0$ and $f_1=0$, the Newton coupling $G(z)$ grows significantly when $r$ decreases. In such a case, there may not exist an inner horizon because $\Delta(r)$ is dominated by the $-2G(z)Mr$ term at small $r$ such that $\Delta(r)$ may only have one root. Similar situations can happen in the case with $e_1=0$ and $f_1<0$. We exhibit these regions of the parameter space in Fig.~\ref{fig:dCdensitynonsingular}, i.e., the region above (below) the red curve on the left (right) panel. The point is that, even if the models in most of the parameter space in Fig.~\ref{fig:dCdensitynonsingular} do not have an inner horizon, the parameter $\epsilon$, which can always be defined for a given parameter set from Eq.~\eqref{defepsilon}, can still capture very well the behavior of the circularity deviation observable $\Delta C$ of the shadow. This can be seen from the black contours in Fig.~\ref{fig:dCdensitynonsingular}.

\subsection{Spin-independent scale identification $(l,p)=(0,1)$} \label{subsec:01}

Finally, we consider the spin-independent scale identification such that the Newton coupling $G(z)$ does not depend on the spin of the black hole. To satisfy the consistency of black hole thermodynamics, i.e., Eq.~\eqref{generalgfunction}, we have $G(r,M,a)=G(Mr)$. Without loss of generality, we choose $(l,p)=(0,1)$ such that
\begin{equation}
z=\frac{1}{Mr}=\frac{1}{G_0M^2r_*}\,.
\end{equation}
The Newton function is again parametrized using Pad\'e expansion
\begin{equation}
G(z)\equiv G_0\left(\frac{1+\sum_{n=1}g_nz^n}{1+\sum_{m=1}h_mz^m}\right)\,.
\label{gexpspinind}
\end{equation}
Expanding the metric component $g_{tt}$ at $r\rightarrow\infty$, we get
\begin{equation}
g_{tt}=-1+\frac{2}{r_*}+\frac{2\left(g_1-h_1\right)}{r_*^2}+\mathcal{O}\left(r_*^{-3}\right)\,.
\end{equation}
Therefore, we need $g_1=h_1$ to satisfy the Solar System constraint \cite{Williams:2004qba}. For simplicity, we assume $g_1=h_1=0$ in this subsection.

\begin{figure*}[htb]
\centering
\includegraphics[width=230pt]{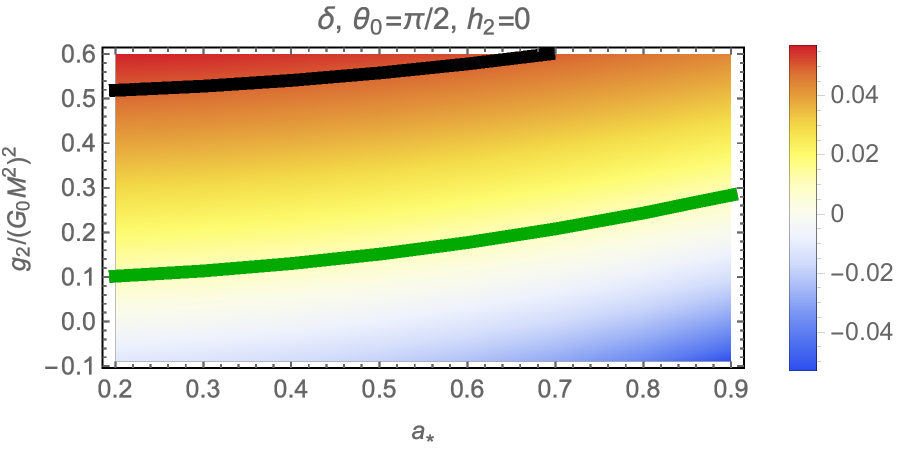}
\includegraphics[width=230pt]{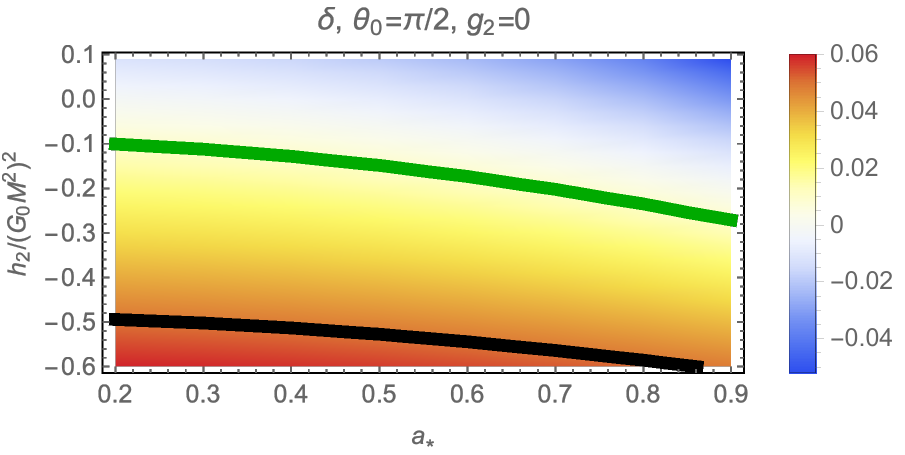}
\caption{The fractional diameter deviation $\delta$ is shown by color in the parameter space of $(a_*,g_2/(G_0M^2)^2)$ (left) and $(a_*,h_2/(G_0M^2)^2)$ (right).}
\label{fig:Rfdensityspinind}
\end{figure*}

\begin{figure*}[htb]
\centering
\includegraphics[width=230pt]{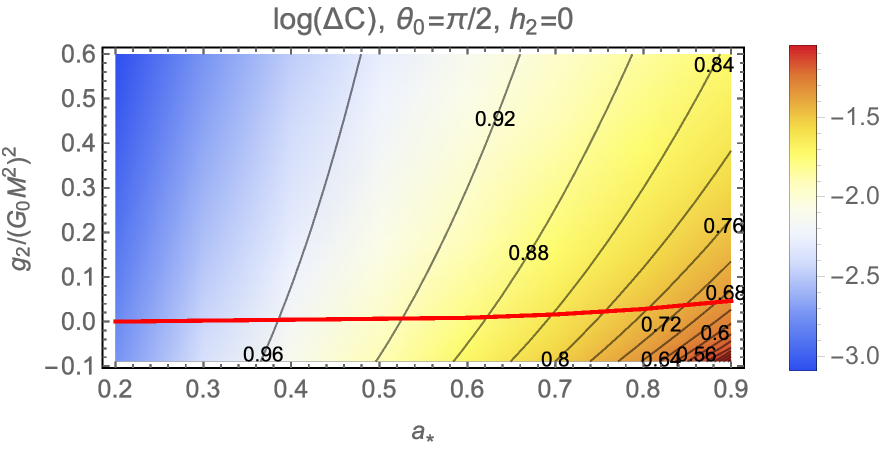}
\includegraphics[width=230pt]{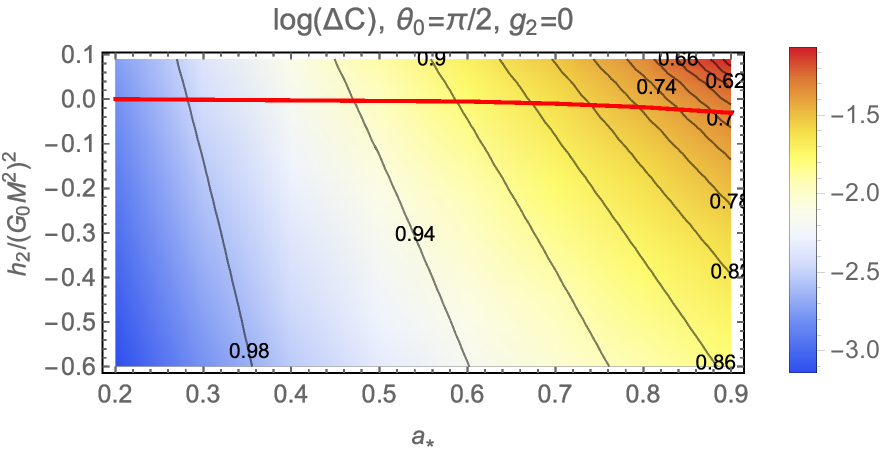}
\caption{The logarithm of the circularity deviation $\Delta C$ is shown by color in the parameter space of $(a_*,g_2/(G_0M^2)^2)$ (left) and $(a_*,h_2/(G_0M^2)^2)$ (right). The black contours represent the values of the extremality deviation $\epsilon$. The parameter space above (below) the red curve on the left (right) panel corresponds to the black hole spacetimes that have only one event horizon.}
\label{fig:dCdensityspinind}
\end{figure*}

In Fig.~\ref{fig:Rfdensityspinind}, we show the fractional diameter deviation $\delta$ in the parameter space of $(a_*,g_2/(G_0M^2)^2)$ and $(a_*,h_2/(G_0M^2)^2)$ on the left and the right panels, respectively. Similar to the previous cases, decreasing the Newton coupling $G(z)$, which can be achieved by decreasing $g_2$ or increasing $h_2$, can shrink the shadow size. The constraints from the Sgr A* shadow size with the Keck and VLTI priors on the mass-to-distance ratio are exhibited.

In Fig.~\ref{fig:dCdensityspinind}, we show the logarithm of $\Delta C$ in the same region of the parameter space as in Fig.~\ref{fig:Rfdensityspinind}. Similar to the case of $(l,p)=(1,4)$, in some regions of the parameter space, there is only one root for $\Delta(r)=0$, and hence only one horizon exists. These regions correspond to the region above (below) the red curves in the left (right) panel. One can see that the extremality deviation $\epsilon$ can nicely capture the behavior of $\Delta C$, even within the parameter space where there is only one event horizon.

\subsection{Hindrance in shrinking shadow size further}\label{subsec:shadowsizelower}

Before ending this section, we would like to point out that for all the models of $(l,p)$ under consideration in this work, shrinking the size of shadows can be achieved by either decreasing the Newton coupling $G(z)$ or increasing the spin $a_*$. This is equivalent to moving on the parameter space toward the extremal configuration, and the shadow critical curves become more distorted, i.e., $\epsilon$ decreases and $\Delta C$ increases. However, if one continues moving in such a direction in the parameter space, one either goes beyond the extremal limit such that the event horizon disappears, or the shadow critical curve does not maintain its D-shaped structure. For instance, in the latter case, the unstable spherical photon orbits may only contribute to an open arc rather than a closed contour on the image plane, and the shadows are not consistent with current observations. This means that in order to have D-shaped shadows, there seems to be a minimum shadow size in the models of different $(l,p)$ under consideration. 

\begin{figure*}[htb]
\centering
\includegraphics[width=300pt]{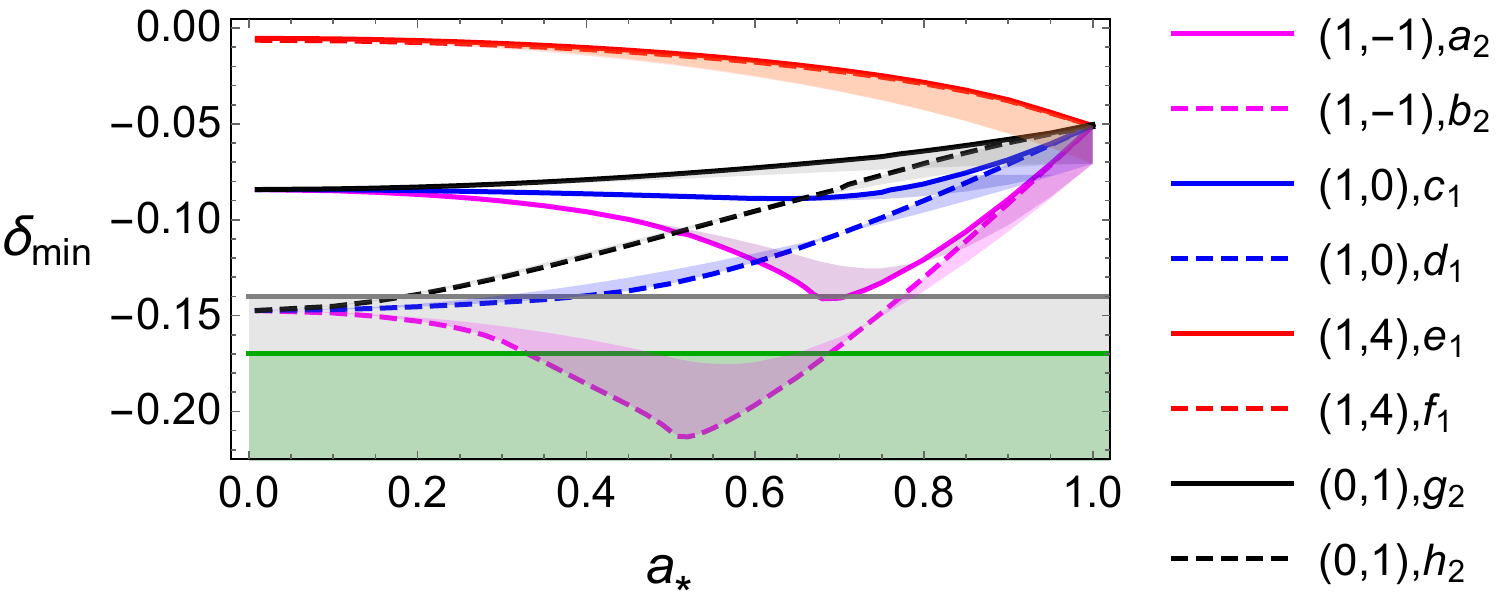}
\caption{The smallest fractional diameter deviation $\delta_\textrm{min}$ of a D-shaped shadow in terms of $a_*$ in models with different $(l,p)$. In the legend, each curve is labeled by its $(l,p)$ and the main varying parameter, with others fixed to be zero. For instance, the solid magenta curve represents $(l,p)=(1,-1)$, with only $a_2$ varying and other coefficients set to zero. For a given $a_*$, the shadow size reaches its minimum value when the main varying parameter is tuned such that the black hole spacetime is near the extremal configuration. The curves represent $\theta_0=\pi/2$, and the shaded regions surrounding the curves represent the results scanned for varied $\theta_0$. For all D-shaped shadows in our model with $|a_*|\le1$, we have $\delta_\textrm{min}\gtrsim -0.21$, i.e., the local minimum of the dashed magenta curve. The horizontal shaded region is beyond the lower bounds obtained from the Keck (gray) and VLTI (green) priors on the mass-to-distance ratio of the Sgr A*.}
\label{fig:sizechange}
\end{figure*}

To demonstrate this in our setup, which has a huge parameter space, we consider different sets of $(l,p)$, then take the spin $a_*$ and one of the coupling coefficients in $G(z)$ as two varying parameters, with other coupling coefficients set to zero. For a given $a_*$, we tune the value of the varying coupling coefficient to identify the smallest fractional diameter deviation $\delta_\textrm{min}$, assuming a D-shaped shadow contour. Even for a given $(l,p)$, when $a_*$ changes, the varying coupling coefficient that corresponds to $\delta_\textrm{min}$ also varies. In Fig.~\ref{fig:sizechange}, we demonstrate the smallest fractional diameter deviation $\delta_\textrm{min}$ of a D-shaped shadow in terms of $a_*$. Each curve represents models with $\theta_0=\pi/2$ and different $(l,p)$ as well as the main varying coupling parameter indicated in the legend. Other coupling parameters are set to zero. In each model with a given $a_*$, {\it the minimum shadow size occurs when the main varying coupling parameter is tuned such that the black hole spacetime is near the extremal configuration.} When $a_*=1$, all models share the same $\delta_\textrm{min}$ with all the coupling parameters vanishing. This corresponds to the extremal configuration of the Kerr black hole. On the other hand, when $a_*=0$, the model of $(l,p)=(1,-1)$ with varying $a_2$ (magenta solid curve), the model of $(l,p)=(1,0)$ with varying $c_1$ (blue solid curve), and the model of $(l,p)=(0,1)$ with varying $g_2$ (black solid curve) degenerate because they both have $G(z)/G_0-1\propto 1/r^2$. Similarly, the model of $(l,p)=(1,0)$ with varying $d_1$ (blue dashed curve), the model of $(l,p)=(0,1)$ with varying $h_2$ (black dashed curve), and the model of $(l,p)=(1,-1)$ with varying $b_2$ (magenta dashed curve) degenerate because they all have $G_0/G(z)-1\propto 1/r^2$. The shaded regions surrounding the curves represent the results scanned for varied $\theta_0$.

The main message in Fig.~\ref{fig:sizechange} is that the resulting smallest fractional diameter deviation $\delta_\textrm{min}$ in different models has a lower bound among all models and all inclinations $\theta_0$ under consideration $\delta_\textrm{min}\gtrsim -0.21$, i.e., the local minimum of the magenta dashed curve.{\footnote{Fixing $2l+p=1$ and choosing $l>1$ could shrink the shadow size further. However, for these cases we have $l+p<0$, hence according to Eq.~\eqref{zgeneralrescale}, the contributions from the running of $G$ are enhanced by the black hole mass. We do not consider such unphysical scenarios.} Since we believe that the four representative models with various $(l,p)$ we have considered in this work should be able to cover the features of a general Newton coupling $G(r,M,a)$ that satisfies the consistency of black hole thermodynamics, such a minimum $\delta_\textrm{min}$ could be a universal lower bound on the shadow size for such a general scenario. In other words, if there exist black holes with D-shaped shadows whose $\delta$ is observed to be smaller than this bound, it implies the violation of any of the assumptions of the setting in our paper, including, of course, the consistency of black hole thermodynamics.

\section{Spinning black holes beyond Kerr bound}\label{sec:superspin}

In Secs.~\ref{subsec:p4} and \ref{subsec:01} in which we considered the models with $(l,p)=(1,4)$ and $(l,p)=(0,1)$, respectively, we pointed out that there are regions of parameter space where $\Delta(r)=0$ only has one root, hence only one horizon exists. One natural question is as follows: in these cases, will the event horizon still exist if one keeps increasing the spin parameter $a_*$ and goes beyond the usual Kerr bound, i.e., $a_*\le1$?

It turns out that it is possible to go beyond the Kerr bound, but the shadow images may develop irregular structures. More explicitly, we find that when $a_*$ is sufficiently large, a branch of stable spherical photon orbits may appear outside the event horizon. This branch of stable spherical photon orbits connects two distinct branches of unstable photon orbits. In this case, the shadow critical curve develops a cuspy structure, as shown in Fig.~\ref{fig:superspinshadow}, i.e., the contours with $a_*=3$ and $a_*=4$.  This can be understood from Fig.~\ref{fig:xieta}, where we show the constants of motion $\xi(r_p)$ and $\eta(r_p)$ given in Eqs.~\eqref{xirp} and \eqref{etarp}, respectively. As we discussed in Sec.~\ref{sec:geodesicshadow}, these constants of motion are functions of the radii $r_p$ of different spherical photon orbits. Note that in this figure, the horizontal axis is expressed by the rescaled radius $r_{p*}\equiv r_p/G_0M$. On the left panel of Fig.~\ref{fig:xieta}, we choose the parameters that correspond to the darker green contour in Fig.~\ref{fig:superspinshadow}, i.e., $a_*=2$. In this case, $\xi(r_p)$ is a monotonically decreasing function in $r_p$, and the orbits that contribute to the contour are all unstable. The shadow critical curve is a smooth and closed contour. On the other hand, on the right panel of Fig.~\ref{fig:xieta}, in which the spin $a_*=4$ is sufficiently large, $\xi(r_p)$ has one local minimum and one local maximum, which are connected by a branch of stable spherical photon orbits (the dashed green segment; see also Eq.~\eqref{rpplp}). This stable branch of orbits connects to two separated branches of unstable orbits on its two ends. There are two distinct unstable orbits, each belonging to the two sides of unstable branches, that contribute to the same point $(\alpha,\beta)$ on the image plane, i.e., the cusps on the contours. Such a discontinuity gives rise to the cuspy structure in the shadow critical curve, and it is a manifestation of the rich spacetime structure in this model, albeit the Liouville integrability of the geodesic dynamics ensured by the Carter constant $\mathcal{K}$ (see Eqs.~\eqref{rthetadot} and \eqref{RTHETASEP}). Black hole models with similar cuspy structures in the shadow critical curves can also be found in Refs.~\cite{Cunha:2017eoe,Wang:2017hjl}.\footnote{Cuspy structure in shadow boundary has also been reported in the models in which two distinct branches of unstable photon orbits transit discontinuously, without resorting to stable spherical photon orbits \cite{Qian:2021qow}.} The model considered in Ref.~\cite{Wang:2017hjl} also has Liouville integrable geodesics. The metric line element of the model considered in Ref.~\cite{Wang:2017hjl} can be expressed in the form of Eq.~\eqref{ksol} with $G(r)=G_0+\bar{\eta}/2Mr^2$ and $\bar\eta$ a non-Kerr parameter. This Newton coupling satisfies thermodynamic consistency only when $\bar{\eta}\propto1/M$. If this choice is made, this falls into the class of solutions discussed in Sec.~\ref{subsec:01}.

Before closing this section, we would like to comment on the physical implications of this type of cuspy shadows and the corresponding black hole spacetimes. First of all, the shadow contours with such a ``circular sector" shape are very different from the usual D-shaped shadows, and can be easily distinguished through observations. Furthermore, from theoretical points of view, the existence of stable spherical photon orbits implies the possibility of accumulating photons (or gravitons) around the black hole for a long time. Such an accumulation may hint toward some sorts of instability of the entire geometry. The relevant questions are whether such systems are truly unstable and, if so, how long the instability timescale is. However, the answer to these questions requires a thorough numerical analysis that takes into account the nonlinear effects of the underlying gravitational theory, as well as the backreactions induced by the accumulating particles. Such an analysis is beyond the scope of this paper. 

\begin{figure*}[h]
\centering
\includegraphics[width=280pt]{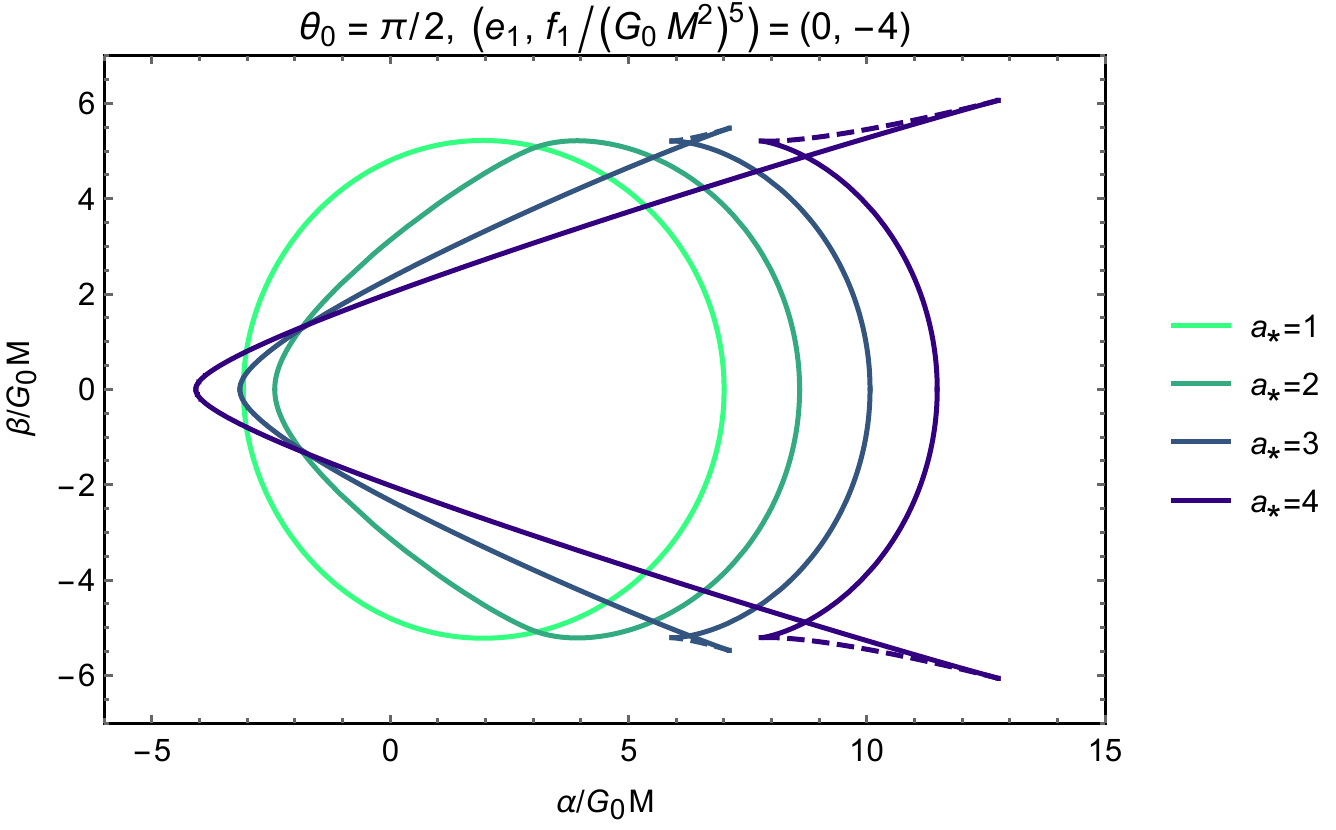}
\caption{The shadow critical curves for black holes with $(l,p)=(1,4)$ discussed in Sec.~\ref{subsec:p4}. In the parameter space where there is only one horizon, one can keep increasing the spin $a_*$. As $a_*$ is sufficiently large, the contours develop irregular deformations, such as cuspy structures at the top and the bottom of the contours, e.g., the contours with $a_*=3$ and $a_*=4$. One can see that two line segments, which are contributed by two branches of unstable photon orbits, extend from the cusp. These two branches of unstable orbits are connected by a branch of stable photon orbits, whose impact parameters are illustrated by the dashed segments (see also Fig.~\ref{fig:xieta}). In this figure, we choose $e_1=0$ and $f_1/(G_0M^2)^5=-4$. The cuspy structure appears when $a_*\gtrsim 2.36$.}
\label{fig:superspinshadow}
\end{figure*}

\begin{figure*}[htb]
\centering
\includegraphics[width=250pt]{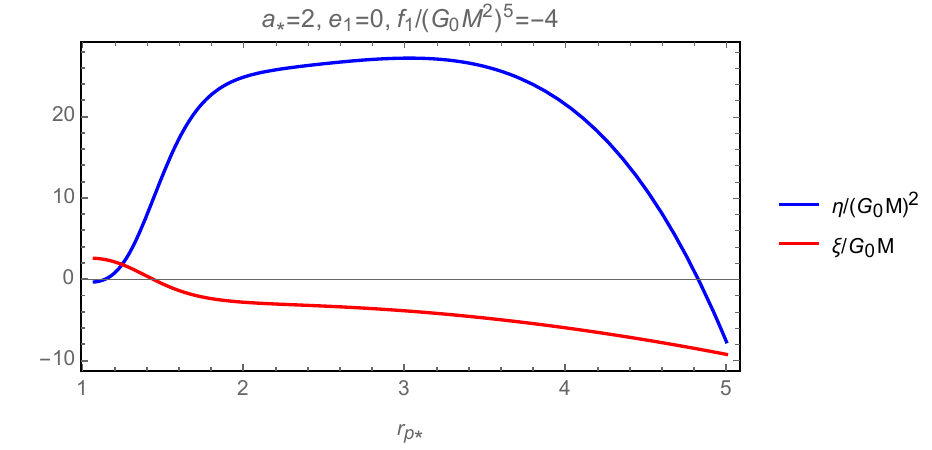}
\includegraphics[width=250pt]{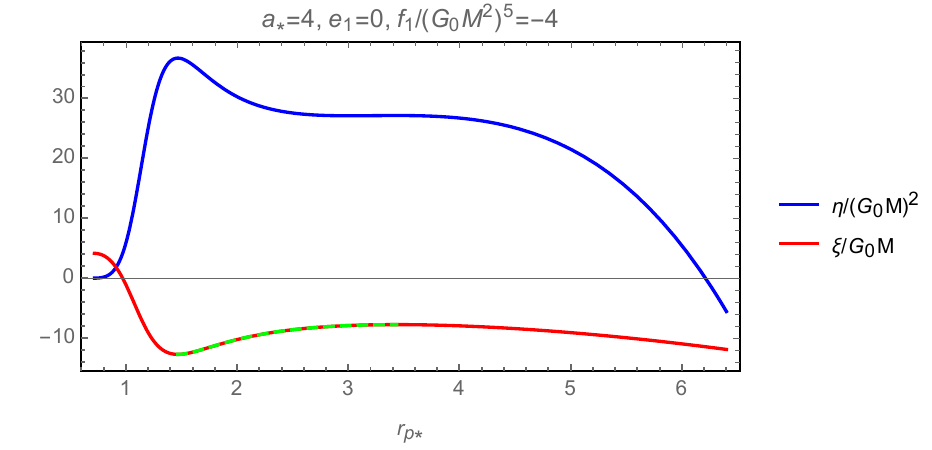}
\caption{Left: for the black holes whose shadow critical curves do not have cuspy structures, the constant of motion $\xi(r_p)$ is a monotonic function of $r_p$. All the spherical photon orbits that contribute to the shadow critical curve are unstable. Right: the cuspy structure on the image plane in Fig.~\ref{fig:superspinshadow} is due to the existence of the local maximum and the minimum in $\xi(r_p)$. In particular, the spherical photon orbits with radii between the two local extrema, i.e., the green dashed line segment, are stable orbits with $d^2\mathcal{R}(r)/dr^2|_{r_p}<0$ (see Eq.~\eqref{rpplp}). The impact parameters of the orbits in this line segment are represented by the dashed segments in Fig.~\ref{fig:superspinshadow}. Note that for $\theta_0=\pi/2$, the spherical photon orbits that contribute to the critical curve are those with radii bounded by the two roots of $\eta(r_p)=0$. The left edge of each panel represents the location of the event horizon, which is very close to the smaller root of $\eta(r_p)=0$.}
\label{fig:xieta}
\end{figure*}

\section{\label{sec:con}Conclusions}

Black hole thermodynamics, an extremely intriguing feature of the Kerr black hole in GR, connects the framework of microscopic physics to the macroscopic entities of classical black holes. In particular, in order to have a well-defined black hole entropy established from the first law, its integrability condition has to be satisfied. This consistency condition of black hole thermodynamics works perfectly for the Kerr black hole, but could be very fragile if one considers non-Kerr geometries. Based on its nature of bridging microscopic physics and macroscopic black hole entities, the consistency of black hole thermodynamics could stand as a guiding principle for constructing effective descriptions of black holes with quantum corrections, or even for building a consistent theory of quantum gravity.

In this paper, we have considered the minimal extension of the Kerr geometry that preserves the thermodynamic consistency, i.e., the line element \eqref{ksol} with a running Newton coupling $G(z)$. This minimal extension respects the circularity of the spacetime, which is a required spacetime symmetry in order to have a well-defined black hole temperature, and also ensures the Liouville integrability of geodesic dynamics. In particular, such an extension can be motivated by the approach of the asymptotic safety based on the FRG, in which the usual coupling constants in classical theory become scale dependent. Modifications of the metric are also considered for the singularity resolution in black holes. The consistency of black hole thermodynamics further restricts the form of the running Newton coupling, i.e., Eq.~\eqref{generalgfunction}. By considering a proper parametrization of the Newton coupling $G(z)$, we have investigated the shadow critical curves, that is, the shadow images contributed by infinitely lensed light rays, cast by this class of black hole models. The main aim of this study is to look for possible implications in terms of shadows if the black hole satisfies the thermodynamic consistency condition. 

Despite the minimal extension, the parametrization of the Newton coupling under consideration gives rise to a huge parameter space in our model. Under the variation of the coupling parameters in $G(z)$, as well as the spin of the black hole, the apparent size and the shape of the shadow critical curves change. This allows us to place constraints on the parameter space using the shadow size measurements available for the Sgr A* images (Figs.~\ref{fig:Rfdensitym1}, \ref{fig:Rfdensitymassind}, \ref{fig:Rfdensitynonsingular}, and \ref{fig:Rfdensityspinind}). In particular, the constraining power using supermassive black hole observations becomes extremely weak for the model in which the black hole singularity can be resolved by some putative quantum effects (Sec.~\ref{subsec:p4}). In addition, we find that the distortion in the shape of shadows, which is quantified by the circularity deviation $\Delta C$, is controlled by not only the spin $a_*$ and the inclination angle $\theta_0$ but also the coupling parameters in $G(z)$. In particular, for the parameter space we have chosen, the behavior of circularity deviation $\Delta C$ can be qualitatively captured by the extremality deviation parameter $\epsilon$ (Figs.~\ref{fig:dCdensitym1}, \ref{fig:dCdensitymassind}, \ref{fig:dCdensitynonsingular}, and \ref{fig:dCdensityspinind}), which is a quantity defined solely on the event horizon (Eq.~\eqref{defepsilon}). The capability of $\epsilon$ to capture the behavior of the circularity deviation $\Delta C$ of the shadow critical curves in such a huge parameter space in our model is totally nontrivial. Moreover, it suggests the possibility of extracting geometrical information of the horizon scale from the shape of the critical curves when the black hole satisfies thermodynamic consistency.    

One particularly crucial result in this paper is the possible lower bound of the apparent shadow size in this model, as we discussed in Sec.~\ref{subsec:shadowsizelower}. In our model, the size of shadow critical curves shrinks as the black hole spins up, or when one decreases the Newton coupling $G(z)$. For a given $a_*$, the shadow contour reaches its minimum size as the black hole approaches the extremal configuration, beyond which either the horizon disappears, or the shadow critical curves cannot maintain the D-shaped structure. We have explored as much as possible the parameter space of our model and all inclinations $\theta_0$ and have identified a lower bound of the fractional diameter deviation $\delta_\textrm{min}\gtrsim -0.21$ for black holes with $|a_*|\leq 1$ (Fig.~\ref{fig:sizechange}). Such a universal lower bound of $\delta$ could be used as an observational indicator to test the consistency of black hole thermodynamics. More explicitly, if there exist black holes whose D-shaped shadows have $\delta$ violating this lower bound, such black holes must violate at least one of the assumptions made in our paper, which include:
\begin{itemize}
\item The setting of the minimal extension, i.e., the line element Eq.~\eqref{ksol} with a running $G(G_0/(r^2+a^2),Mr)$. 
\item The parametrization for $G(G_0/(r^2+a^2),Mr)$. 
\item The consistency of black hole thermodynamics.
\end{itemize}

Another interesting result we obtain in this paper is the possibility of violating the Kerr bound $|a_*|\leq1$ in this model, as we discussed in Sec.~\ref{sec:superspin}. This is because in some regions of the parameter space, the black holes have only one event horizon, even if they have nonzero spin (Figs.~\ref{fig:dCdensitynonsingular} and \ref{fig:dCdensityspinind}). However, for a sufficiently large spin $a_*$, a new branch of stable spherical photon orbits may appear outside the event horizon, and the shadow critical curves develop cuspy structures (Fig.~\ref{fig:superspinshadow}). Such systems may suffer from issues of instability due to the existence of stable spherical photon orbits. In any case, this result is a manifestation of the rich spacetime structure inherent in this model, even if the model is based on a minimal extension from the Kerr spacetime with integrable geodesic dynamics.

We would like to emphasize that although the spacetime metric \eqref{ksol} with a running Newton coupling preserves the Liouville integrability of geodesic dynamics, which implies that the model is a subclass of the general metric parametrizations considered in Refs.~\cite{Johannsen:2013szh,Papadopoulos:2018nvd}, the restriction on the form of the running Newton coupling required by the consistency of thermodynamics significantly reduces the arbitrariness of the metric parametrizations. Therefore, although the parameter space of our model is still large, the model cannot generate arbitrary shadows, and that is why we have been able to identify common shadow features (e.g., the behavior of $\epsilon$ and the universal lower bound of $\delta$) shared in such a huge parameter space. These shadow features may not exist anymore when going beyond the subclass our model belongs to, where the consistency of thermodynamics may be broken.

One possible extension of this work is to go beyond the minimal setting in Eq.~\eqref{ksol} by considering, for instance, the most general Liouville integrable spacetime \cite{Benenti:1979erw,Papadopoulos:2018nvd}. One can even relax the assumption of Liouville integrability and consider a more general spacetime parametrization \cite{Konoplya:2016jvv}. It is highly desirable to identify the conditions for thermodynamic consistency in these general spacetime settings, and to look for possible observational implications. We will leave these interesting issues to future work.

\section*{Acknowledgements}
We would like to thank Li-Ming Cao for valuable discussions. C.-Y. C. is supported by the Special Postdoctoral Researcher (SPDR) Program at RIKEN and RIKEN Incentive Research Grant (Shoreikadai) 2025. C.-M. C. and N. O. would like to thank Asia Pacific Center for Theoretical Physics for the hospitality
where this work was completed. The work of C.-M. C. was supported by the National Science and Technology Council of the R.O.C. (Taiwan) under the grants NSTC 114-2112-M-008-010.


\bibliographystyle{utphys}
\bibliography{bib}

\end{document}